\documentclass[aps,prd,a4paper,twocolumn]{revtex4-2}
\usepackage{header_revtex}

\begin{document}

\date{\today}
\title{Phenomenological quark-hadron equations of state with first-order phase transitions for astrophysical applications}

\author{Niels-Uwe F. Bastian}
\affiliation{Institute for Theoretical Physics, University of Wroclaw, pl. M. Borna 9, 50-204 Wroclaw, Poland}

\begin{abstract}
In the current work an \glsentrylong{eos} model with a first-order phase transition for astrophysical applications is presented.
The model is based on a two-phase approach for quark-hadron phase transitions, which leads by construction to a first-order phase transition.
The resulting model has already been successfully used in several astrophysical applications, such as cold neutron stars, core-collapse supernova explosions and binary neutron star mergers.
Main goal of this work is to present the details of the model, discuss certain features and eventually publish it in a tabulated form for further use.
\end{abstract}


\maketitle

\section{Introduction}
\label{sec:intro}
The investigation of the \gls{eos} of strongly interacting matter is an ongoing problem of nuclear and high-energy physics.
Direct approaches to solve the underlying theory of \gls{qcd} are only accessible at high temperatures and vanishing densities or at asymptotically high densities.
For the region, which is relevant in astrophysics, mainly effective and phenomenological approaches are in use (e.g. see reviews \cite{Fischer:2017zcr,Oertel:2016bki} for further reference).

Of particular interest is the possible transition from ordinary hadronic matter at low densities/temperatures to a phase of deconfined quarks.
Numerical results from \gls{lqcd} predict a crossover transition with a pseudo-critical temperature of $T(\mu_\mathrm B = 0) = \np{156.5}\,\np[MeV]{\pm1.5}$ \citep{Bazavov:2018mes}.
However, these  calculations are limited to small baryon chemical potentials and can not reach densities relevant for astrophysics.
At asymptotically high densities \gls{pqcd} can be applied, which predicts a phase of deconfined quark matter \citep{Kraemmer:2003gd,Kurkela:2014vha}.
Unfortunately, it is not possible to reach densities, where the transition from hadronic to quark matter occurs, and therefore both the position and order of this transition at low temperatures are currently speculative.
The two possible scenarios are the existence of at least one \gls{cep}, changing the crossover transition into a first-order transition, and the absence of any \gls{cep}, which would result in the crossover transition to span the entire phase diagram.

In the presented model, the existence of a first-order phase transition at high baryon densities is assumed as a working hypothesis, in order to explore possible implications in astrophysics, which might lead to measurable signals.
Contrary to the original publication of the model in \cite{Kaltenborn:2017hus}, where neutron star configurations are studied, here the applicability is extended to such dynamic phenomena as \gls{ccsn} and \gls{bnsm}, which reach not only high densities, but also high temperatures.
This extension broadens the spectrum of predictable signals from mass--radius relations and tidal deformability (of neutron stars) to, e.g., gravitational waves, neutrino signals and possibly electromagnetic counter parts.
Additionally, the hypothetical case of multiple \glspl{cep} is covered, due to the possibility to probe the existence of a first-order phase transition at various temperatures.

The material presented here was so far successfully applied to \gls{bnsm} in \cite{Bauswein:2018bma}, suggesting a possible signal in gravitational waves, which would unambiguously identify the existence of a sudden softening in the \gls{eos}, which is characteristic of a first-order phase transition.
Furthermore, it was applied to \gls{ccsn} simulations in \cite{Fischer:2017lag}, predicting a second shock wave, which leads to the successful explosion of a $\np[M_\odot]{50}$ progenitor star and a measurable neutrino signal, originating from the phase transition.
The additional shock wave in the \gls{ccsn} leads to an altered result of nucleosynthesis analyses, bringing back supernovae as source of $r$-process elements, as shown in \cite{Fischer:2020xjl}.

The document is structured in the following way: First of all, \cref{sec:model} presents the model in its current state, including the description of each component and the realization of phase transition.
Afterwards, in \cref{sec:results} the created parameter sets and their properties are shown.
Finally, in \cref{sec:discussion} a discussion is laid out about the consequences of certain aspects of the model and possible alternative approaches.

Note, that in this publication natural units $\hbar = c = k_\mathrm B = 1$ and the unity volume $V = 1$ is applied.

\section{Hybrid equation of state model}
\label{sec:model}
The particles involved in astrophysical systems can be grouped into 5 classes.
Due to the different nature of their thermodynamic interactions it is reasonable to assume, that mixed terms in the thermodynamic potential are small and their contributions can be addressed separately:
\begin{multline}
    \Omega = \Omega_\mathrm{nucleons} + \Omega_\mathrm{nuclei} + \Omega_\mathrm{quarks}\\
    + \Omega_\mathrm{leptons} + \Omega_\mathrm{photons}\,.
\end{multline}
Since photons do not carry any charge, they are only included as thermal excitation at high temperatures and do not affect the phase transitions.
Included leptons, such as electrons and neutrinos, are fully degenerate at our densities and can be described as ideal gas.
They carry electric and leptonic charge and are (dependent on the system) in chemical equilibrium with the strongly-interacting particles.
The separation of the strongly-interacting part into quarks and hadrons is an assumption of the presented model, which is commonly used in astrophysics (so called two-phase approach).

In \cref{sec:model:rdf} the \gls{rdf} formalism is derived, which is used to describe homogeneous quark and hadron matter in a mean field approximation.
Afterwards, the particular models used for  $\Omega_\mathrm{nucleons}$ (\cref{sec:model:hadrons}) and $\Omega_\mathrm{quarks}$ (\cref{sec:model:quarks}) are shown.
\Cref{sec:model:nuclei} presents the model which is used for nuclear cluster formation $\Omega_\mathrm{nuclei}$ and \cref{sec:model:leptons} explains the inclusion of leptons $\Omega_\mathrm{leptons}$.

\subsection{\Glsentrylong{rdf} derived from field theory}
\label{sec:model:rdf}

The \gls{rdf} approach as introduced in \cite{Kaltenborn:2017hus} is a framework capable of dealing with very complex interaction contributions, e.g., confinement.
Here the derivation is done in a more thorough and complete manner, including isovector couplings, which have major effects on the isospin asymmetric phase diagram, relevant for astrophysics.

The derivation is done self-consistently from the path-integral formalism, based on an effective Lagrangian of low-energy \gls{qcd} to obtain the partition function $\mathcal Z$ and hence the thermodynamic potential
\begin{align}
	\Omega &= -T \ln \mathcal Z\,.
\end{align}
Analogous to the treatment of the Walecka model of nuclear matter in \cite{Kapusta:1989tk}, the partition function takes the form
\begin{align}\label{eq:dft:Z}
	\mathcal{Z} &= \int \pathint{\bar q}\pathint{q} \exp\left\{\int\mathrm d^4 x
	\left[\mathcal{L}_\mathrm{eff} + \bar{q}\gamma^0 \hat{\mu} q\right]\right\}~,
\end{align}
where in the case of a two-flavour quark model
\begin{align}
	q=\left(\begin{array}{c}q_\mathrm u\\q_\mathrm d\end{array}\right),
\end{align}
and $\hat{\mu} = \operatorname{diag}(\mu_\mathrm u,\mu_\mathrm d)$ is the diagonal matrix of the chemical potentials conjugate to the conserved numbers of corresponding quarks.
The effective Lagrangian density is given by
\begin{align}
	\label{eq:dft:Leff}
	\mathcal{L}_\mathrm{eff} &= \mathcal{L}_\mathrm{free} - U\,,\\
	\mathcal L_\mathrm{free} &= \bar q \left(\imath\gamma^\mu \partial_\mu - \hat m\right) q~,
\end{align}
where $\hat{m} = \operatorname{diag}(m_\mathrm u,m_\mathrm d)$ is the matrix of current quark masses.
The interaction is given by the potential energy density $U = U(\bar qq, \bar q\vec\tau q, \bar q\gamma^0q, \bar q\vec\tau\gamma^0q)$ which in general is a non-linear functional  of the field representations of the scalar, vector and corresponding isovector quark currents.
Note here, that I restrict myself to quark fields out of readability of the manuscript - the formalism can be written down for any fermionic many-body system, which needs to be treated in a relativistic scheme by exchanging the respective quark fields $q$ and $\bar q$ by, e.g., nucleonic fields $\psi$ and $\bar\psi$.

In the isotropic case, the vector four-current reduces to its zeroth component and the gradients of the fields vanish, so that $\gamma^\mu \partial_\mu = \gamma^0\partial_0$.
Due to charge conservation, only the third component of any isovector field would remain, so that only the third component of the isospin--vector $\vec\tau = (\tau_1, \tau_2, \tau_3)^\mathrm T$ remains, which is here short-handed written as $\tau = \tau_3$.

To achieve a quasi-particle representation, the potential energy density shall depend linearly on the Dirac spinor bilinears representing the relevant currents of the system.
The linearization of the interaction is facilitated by a Taylor expansion around the corresponding expectation values
\begin{subequations}
\label[pluralequation]{eq:rdft:densities}
\begin{align}
	\label{eq:rdft:ns} \left<\bar{q}q\right> &= \ns = \sum_i \ns_i = -\sum_i T\frac{\partial }{\partial m_i} \ln \mathcal{Z}\,,\\
	\label{eq:rdft:nsi} \left<\bar{q}\tau q\right> &= \nsi = \sum_i \tau_i \ns_i = -\sum_i \tau_i T\frac{\partial }{\partial m_i} \ln \mathcal{Z}\,,\\
	\label{eq:rdft:nv} \left<\bar{q}\gamma^0q\right> &= \nv =\sum_i \nv_i = \sum_i T \frac{\partial }{\partial \mu_i} \ln \mathcal{Z}\,,\\
	\label{eq:rdft:nvi} \left<\bar{q}\tau \gamma^0q\right> &= \nvi =\sum_i \tau_i \nv_i = \sum_i \tau_i T \frac{\partial }{\partial \mu_i} \ln \mathcal{Z}\,,
\end{align}
\end{subequations}
of the scalar density $\ns$, the vector density $\nv$, the scalar--isovector density $\nsi$ and the vector--isovector density $\nvi$, respectively.
Here $\tau_i$ is the isospin quantum number of the particle species $i$.
This expansion results in
\begin{multline}
	U =
	\bar U
	+ (\bar{q}q - \ns)\SigmaS
	+ (\bar{q}\tau q - \nsi)\SigmaSI\\
	+ (\bar{q}\gamma^0q - \nv) \SigmaV
	+ (\bar{q}\tau\gamma^0q -  \nvi) \SigmaVI +\ldots~,
\end{multline}
where the notation
\begin{subequations}
\begin{align}
	\SigmaS &= \left.\frac{\partial U}{\partial (\bar{q}q)} \right|_{*} = \frac{\partial \bar U}{\partial \ns} \,,\\
	\SigmaSI &= \left.\frac{\partial U}{\partial (\bar{q}\vec\tau q)} \right|_{*} = \frac{\partial \bar U}{\partial \nsi} \,,\\
	\SigmaV &= \left.\frac{\partial U}{\partial (\bar{q}\gamma^0q)} \right|_{*} = \frac{\partial \bar U}{\partial \nv}\,,\\
	\SigmaVI &= \left.\frac{\partial U}{\partial (\bar{q}\vec\tau\gamma^0q)} \right|_{*} = \frac{\partial \bar U}{\partial \nvi}\,,
\end{align}
\end{subequations}
respectively for the different self energies, is introduced.
The derivatives (marked by the asterisk $*$) are taken at the expectation values of the field bilinears $\bar{q}q = \ns$, $\bar{q}\tau q = \nsi$, $\bar{q}\gamma^0q = \nv$, and $\bar{q}\tau\gamma^0q = \nvi$.
The newly defined $\bar U = U(\ns, \nsi, \nv, \nvi)$ is the potential energy density at the expectation values.
The expansion is truncated at the second term, assuming the fluctuations around the expectation values of the fields are small.
Applying this quasi-particle approximation to the effective Lagrangian of \cref{eq:dft:Leff} and reordering the terms results in
\begin{align}
	\mathcal L_\mathrm{eff} &= \mathcal{L}_{\textrm{qu}} - \Theta(\ns, \nsi, \nv, \nvi)\,,
\end{align}
with the quasi-particle contribution
\begin{multline}
	\mathcal L_\mathrm{qu} = \bar q \big(\gamma^0 \left(\imath\partial_0  + \SigmaV + \tau\SigmaVI\right)\\
	- \left(\hat m + \SigmaS + \tau\SigmaSI\right)\big) q\,,
\end{multline}
and the effective potential energy density
\begin{align}
	\Theta &= \bar U - \SigmaS \ns - \SigmaSI \nsi - \SigmaV \nv - \SigmaVI \nvi~.
\end{align}
Now, the partition function from \cref{eq:dft:Z} takes the form
\begin{align}\label{eq:dft:Zeff}
	\mathcal Z &= \!\!\int \mathcal D\bar q\mathcal Dq \exp\left[\mathcal{S}_\mathrm{qu}[\bar q,q] - \frac 1T \Theta(\ns,\nsi,\nv,\nvi)\right]\!\!
\end{align}
where the quasi-particle action in Fourier-Matsubara representation is given by \cite{Kapusta:1989tk}
\begin{align}
	\mathcal S_\mathrm{qu}[\bar{q},q] &= \frac 1T\sum_n\sum_{\vec{p}} \bar{q}~ G^{-1}(\omega_n,\vec{p})~ q~,\\
	G^{-1}(\omega_n,\vec{p}) &= \gamma^0(-i\omega_n + \hat{\tilde\mu}) -\vec{\gamma}\cdot\vec{p} - \hat{M}~,
\end{align}
with the Dirac effective masses $M_i = m_i + \SigmaS + \tau_i\SigmaSI$ and the renormalized chemical potential $\tilde\mu_i = \mu_i - \SigmaV - \tau_i\SigmaVI$.
Note here that the hat notation of $\hat{\tilde\mu}$ and $\hat{M}$ stands again for diagonal matrix over all species.
For convenience the short-hand notations
\begin{align}
	S_i &= \SigmaS + \tau_i\SigmaSI\,,\\
	V_i &= \SigmaV + \tau_i\SigmaVI\,,
\end{align}
are introduced, defining the scalar shift $S_i$ and the vector shift $V_i$, respectively for each particle species $i$.
The functional integral can be performed in this quasi-particle approximation with the result
\begin{align}\label{eq:dft:Zqu}
	\mathcal{Z}_\mathrm{qu} = \int \mathcal{D}\bar{q}\,\mathcal{D}q\, \exp\left\{\mathcal{S}_\mathrm{qu}[\bar{q},q]\right\}=\det[\frac 1T G^{-1}]~,
\end{align}
where the determinant operation acts in momentum--frequency space as well as on the Dirac, flavour, and colour indices.
Using the identity $\ln \det A = \Tr \ln A$ and the representation of the gamma matrices, one obtains for the thermodynamic potential (for details see, e.g., \cite{Kapusta:1989tk})
\begin{align}
\begin{split}
	\Omega_\mathrm{qu} &= -T \ln \mathcal{Z}_\mathrm{qu} = - T \Tr \ln [\frac 1T G^{-1}]\\
	&= \sum_i g_i \int \frac{\mathrm d^3 p}{(2\pi)^3} T \left(\ln \left[1 + \ee^{-\frac 1T (E^*_i-\tilde\mu_i)} \right]\right.\\
	&\qquad\qquad\qquad\qquad\left.+ \ln \left[1+\ee^{-\frac 1T(E^*_i+\tilde\mu_i)} \right]\right)\,,
\end{split}
\end{align}
where the so-called ``no sea'' approximation (as is customary in the Walecka model) is tacitly used by removing the vacuum energy term which corresponds to the phase space integral over the kinetic one-particle energy $E^*_i=\sqrt{p^2+M_i^2}$.

\subsubsection*{Self-consistency}

In order to evaluate the thermodynamics of the \gls{rdf} approach, one has to solve a self-consistency problem, since the thermodynamic potential is a functional of the scalar and vector densities, which themselves are defined as derivatives of the thermodynamic potential by \cref{eq:rdft:densities}.
The thermodynamic potential takes now the form
\begin{multline}
	\Omega = -\sum_i g_i \int \frac{\mathrm d^3 p}{(2\pi)^3} T \left(\ln \left[1 + \ee^{-\frac 1T (E^*_i-\tilde\mu_i)} \right]\right.\\
	\left.+ \ln \left[1+\ee^{-\frac 1T(E^*_i+\tilde\mu_i)} \right]\right) + \Theta\,,
\end{multline}
while its derivatives are
\begin{align}
	\ns_i (T,\{\mu_j\}) &= \left(\frac{\partial \Omega (T,\{\mu_j\})}{\partial m_i}\right)_{T,\{\mu_j\},\{m_{j\neq i}\}}\notag\\
    &= \int \frac{\mathrm d^3 p}{(2\pi)^3} \frac{M_i}{\sqrt{p^2 + M_i^2}} \left(f_i + \bar f_i\right)\,,\\
	\nv_i (T,\{\mu_j\}) &= - \left(\frac{\partial \Omega (T,\{\mu_j\})}{\partial \mu_i}\right)_{T,\{\mu_{j\neq i}\}}\notag\\
	&= \int \frac{\mathrm d^3 p}{(2\pi)^3} \left(f_i - \bar f_i\right)\,,
\end{align}
with the Fermi distributions for particles and antiparticles
\begin{align}
	f_i &= \frac{1}{\ee^{(\sqrt{p^2 + M_i^2}-\tilde\mu_i)/T}+1}\,,\\
	\bar f_i &= \frac{1}{\ee^{(\sqrt{p^2 + M_i^2}+\tilde\mu_i)/T}+1}\,.
\end{align}
Once this set of equations is solved, one can compute all thermodynamic quantities explicitly.

Based on this formalism, the specific interaction potential for the hadronic and quark models are introduced and the resulting self energies $S_i$ and $V_i$ are discussed in the next section.

\subsection[Hadron Model]{Hadron model -- Walecka model in the \glsentryshort{rdf} formulation}
\label{sec:model:hadrons}
Within this section it will be shown how the \gls{dd2} \gls{eos} can be written in the \gls{rdf} formalism.
While the \gls{rmf} approach with density-dependent couplings was already depicted in \cite{Typel:1999yq}, the current parametrization can be found in \cite{Typel:2009sy}.
In this work, \gls{dd2} is the hadronic \gls{eos} of choice, because it is well established in astrophysics and provides an excellent reproduction of nuclear properties as well as withstands all constraints, which are important in astrophysical applications, except the flow-constraint, that has been corrected in the \gls{dd2f} parametrization (see next section).

Both \gls{dd2} and \gls{dd2f} model can be expressed in the \gls{rdf} formalism with the potential
\begin{align}
	U &= -\frac 12 \frac{\Gamma_\sigma^2}{m_\sigma^2} \ns^2
	+ \frac 12 \frac{\Gamma_\omega^2}{m_\omega^2} \nv^2
	+ \frac 12 \frac{\Gamma_\rho^2}{m_\rho^2} \nvi^2~,
\end{align}
with the density-dependent coupling parameters $\Gamma_{\{\sigma,\omega,\rho\}}$, the meson masses $m_{\{\sigma,\omega,\rho\}}$ and the corresponding densities.
It results in the scalar self energy
\begin{align}
	S &= - \frac{\Gamma_\sigma^2}{m_\sigma^2} \ns~,
\end{align}
and the species-dependent vector self energy
\begin{multline}
	V_i = \frac{\Gamma_\omega^2}{m_\omega^2} \nv
	+ \tau_i\frac{\Gamma_\rho^2}{m_\rho^2} \nvi\\
	- \frac{\Gamma_\sigma \Gamma_\sigma' }{m_\sigma^2} \ns^2
	+ \frac{\Gamma_\omega \Gamma_\omega' }{m_\omega^2} \nv^2
	+ \frac{\Gamma_\rho \Gamma_\rho' }{m_\rho^2} \nvi^2~.
\end{multline}
Using these quantities, one can formulate all thermodynamic observables, as described in \cref{sec:model:rdf}.

The form of the coupling terms is not altered in this representation and can be used from the original work.

\subsubsection{Corrections due to flow constraint -- \glsentryname{dd2f}}
With regard to the flow constraint \citep{Danielewicz:2002pu}, the behaviour at supersaturation densities was altered in \cite{Alvarez-Castillo:2016oln} by redefining the coupling parameters as
\begin{align}
	\Gamma_i^\mathrm{DD2f} &= F_i \Gamma_i^\mathrm{DD2}\,,
\end{align}
with $F_i$ as new density-dependent function
\begin{align}
	F_i &= \begin{cases}
		\frac{1 + k (1 + p) y^m}{1 + k (1 - p) y^m}	& \text{for}\quad i = \sigma\\
		\frac{1 + k (1 - p) y^m}{1 + k (1 + p) y^m}	& \text{for}\quad i = \omega\\
		1	& \text{for}\quad i = \rho\\
	\end{cases}\,,
\end{align}
and
\begin{align}
	y &= \begin{cases}
		n/n^\mathrm{ref} - 1	& \forall\; n > n^\mathrm{ref}\\
		0	                    & \forall\; n \le n^\mathrm{ref}
	\end{cases}\,.
\end{align}
The occurring parameters are set to $k = \np{0.04}$, $p = \np{0.07}$, and $m = \np{2.25}$.
The reference density $n^\mathrm{ref} = n^\mathrm{sat} = \np[fm^{-3}]{0.149}$ is again set to the model's saturation density.
One can see, that for $n\leq n^\mathrm{ref} \rightarrow F_i = 1$ it returns to original \gls{dd2}, so the change does not touch the behaviour below saturation density, where \gls{dd2} is already a very sophisticated model.
The coupling to the $\rho$ field is not altered at all, resulting in an unchanged asymmetry behaviour.

\subsubsection{Nuclear properties}

Both \gls{dd2} and \gls{dd2f} models give the same nuclear properties.
The nuclear saturation density is $n^\mathrm{sat} = \np[fm^{-3}]{0.149}$ and the corresponding binding energy is $E_\mathrm{B} = \np[MeV]{16.02}$.
The incompressibility and symmetry energy at saturation can be found as $K = \np[MeV]{255.3}$ and $E_\mathrm{Sym} = \np[MeV]{31.674}$.
Furthermore, they feature a critical endpoint of the liquid-gas phase transition at baryon density  $n^\mathrm{crit} = \np[fm^{-3}]{0.0448}$ and temperature $T^\mathrm{crit} = \np[MeV]{13.71}$.
Important information for astrophysics are the maximal mass for neutron stars $M_\mathrm{max}^\mathrm{DD2} = \np[M_\odot]{2.44}$, $M_\mathrm{max}^\mathrm{DD2F} = \np[M_\odot]{2.08}$  and the direct-Urca threshold density $n_\mathrm{DU} = \infty$.

For a recent overview on this class of hadronic \gls{eos} models and how they are adjusted to nuclear observables please see \cite{Typel:2020ozc}.

\subsection[Quark Model]{Quark model -- confinement as density functional}
\label{sec:model:quarks}

The nature of quarks and gluons, which are strongly interacting particles, is yet little understood.
The \gls{qcd} as underlying theory of strong interaction can only be solved numerically at vanishing baryon-chemical potential, or with perturbative methods in the asymptotic limit of infinite temperature or density.
For the region of dense, yet non-perturbative matter, recently most of the approaches employ \gls{njl}-type models for the description of chiral symmetry breaking (restoration) but lack the deconfinement.
The aim of the current work is to describe all phenomena, known for quark interaction, in a consistent way.
To this end I adopt the following density functional for the interaction energy:
\begin{multline}\label{eq:dft:Ufull}
	U(\ns,\nv,\nvi)
	= C \ns^{4/3}
	+ D(\nv)\ns^{2/3}\\
	+ \omega_4 \nv^2 + \frac{\omega_8}{1 + \omega_8' \nv^2} \nv^4
	+ \rho_4 \nvi^2\,.
\end{multline}
Because of the number of terms occurring in \cref{eq:dft:Ufull}, each representing a physical effect, I will go through them in several subsections.

\subsubsection{Quark confinement}
\label{sec:model:quarks:confinement}

\begin{figure}
	\centering
	\includegraphics[scale=\agrscale]{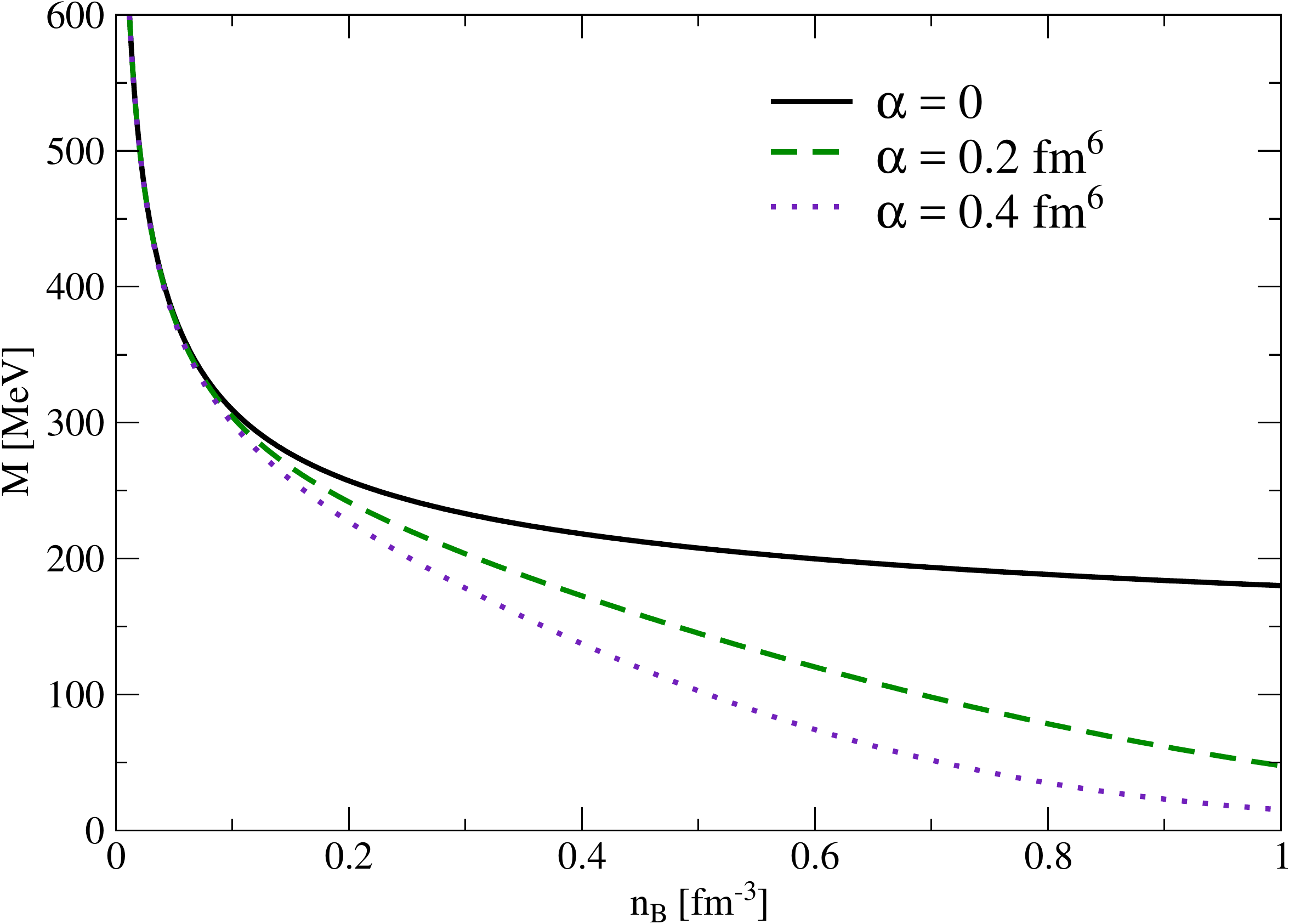}
	\caption{%
	Effective quark mass due to confinement for different colour screening parameters $\alpha$, see \cref{eq:dft:sf_volfrac}. The effective confinement manifests itself by a divergence of the quasi-particle mass at low densities.
	Here $\sqrt{D_0} = \np[MeV]{240}$ and $C = 0$, see \cref{eq:dft:Usfn}.
	}
	\label{fig:dft:sfm_M_nb}
\end{figure}

The first terms in \cref{eq:dft:Ufull}
\begin{align}\label{eq:dft:Usfn}
	U_\mathrm{SFM}(\ns,\nv)
	&= C \ns^{4/3}
	+ D(\nv)\ns^{2/3}~,
\end{align}
represent the so-called \gls{sfm} as introduced in \cite{Kaltenborn:2017hus}.
It captures aspects of (quark) confinement through its resulting density-dependent scalar self energy contribution to the effective quark mass $M$:
\begin{align}
	S_\mathrm{SFM} &= \frac{4}{3} C \ns^{1/3} + \frac{2}{3}D(\nv)\ns^{-1/3}\,.
\end{align}
Here the first term is motivated by the Coulomb-like one-gluon exchange, while the second term represents the linear string potential between quarks.
The effective mass diverges for densities approaching zero, see \cref{fig:dft:sfm_M_nb}, and thus suppresses the occurrence of the quasi-particle excitations corresponding to these degrees of freedom.
For colour neutral hadrons this divergence of the self energy is entirely compensated by that of the confining interaction in the equation of motion \citep{Glozman:2008fk}.
For quark matter in compact stars, such a mechanism has been used in \cite{Li:2015ida}.
Note that in its non-relativistic formulation with energy shifts \citep{Ropke:1986qs}, the \gls{sfm} has already been applied successfully to describe massive hybrid stars with quark-matter cores \citep{Blaschke:1989nn}.

The \gls{sfm} modification takes into account the occupation of the surrounding medium by colour fields, which leads to an effective reduction of the in-medium string tension.
This is achieved by multiplying the vacuum string tension parameter $D_0$ with the available volume fraction $\Phi(\nv)$:
\begin{align}
\label{Dv}
	D(n_\mathrm v) &= D_0 \Phi(\nv)~.
\end{align} 
This reduction of the string tension is understood as a consequence of a modification of the pressure on the colour field lines by the dual-Meissner effect since the reduction of the available volume corresponds to a reduction of the non-perturbative dual superconductor \gls{qcd} vacuum that determines the strength of the confining potential between the quarks.

A standard ansatz for the volume fraction is the Gaussian approach
\begin{align}
	\label{eq:dft:sf_volfrac}
	\Phi(n_\mathrm v) &= \begin{cases}
		\exp\left[-\alpha (\nv - n^\mathrm{ref})^2\right]&	\nv > n^\mathrm{ref}\\
		1&	\nv \le n^\mathrm{ref}
	\end{cases}\,
\end{align}
dependent on the vector density $\nv$, a volume fraction parameter $\alpha$, which scales the reduction and a reference density $n^\mathrm{ref}$ from which the effect starts.
The reference density is set to zero in the presented models.

\subsubsection{Vector repulsion}
\label{sec:model:quarks:vector}
The following contribution in \cref{eq:dft:Ufull}, as it was already introduced in \cite{Kaltenborn:2017hus}
\begin{align}\label{eq:dft:Uomega}
	U_\mathrm{v}(\nv)
	&= \omega_4 \nv^2
	+ \frac{\omega_8}{1 + \omega_8' \nv^2} \nv^4
\end{align}
stands for the repulsion stemming from a four-fermion interaction in the Dirac vector channel and a higher-order (8-fermion) repulsive interaction in the vector channel.
Such higher-order vector mean fields have already been considered in the description of nuclear matter (see, e.g. \cite{Serot:1997xg}), and it is therefore natural to invoke them also in the description at the quark level.
The higher-order quark interactions have been introduced in \cite{Benic:2014iaa} for the description of hybrid stars in order to provide a sufficient stiffening at high densities required to fulfil the $2M_\odot$ mass constraint from the precise mass measurement of \cite{Demorest:2010bx} and \cite{Antoniadis:2013pzd}.
This allows one to obtain a separate third family of high-mass hybrid stars \citep{Benic:2014jia}.
The denominator $(1 + \omega_8' \nv^2)^{-1}$ in the higher-order term compensates its asymptotic behaviour to ensure the speed of sound $c_\mathrm s=\sqrt{\partial P/\partial \varepsilon}$ does not exceed the speed of light.
It can be seen as a form factor of the 8-fermion interaction vertex.

\subsubsection{Isospin mean field}
\label{sec:model:quarks:isospin}
The last contribution in \cref{eq:dft:Ufull}
\begin{align}
	U_\mathrm{vi}(\nvi) &= \rho_4 \nvi^2\,,
\end{align}
is representing a vector--isovector interaction, comparable to the $\rho$-meson interaction in Walecka-type models.
An isovector mean field is usually not considered in quark models, since the symmetry energy of quark matter is not known.
On the other hand, such models are often designed to fit only neutron stars, in which case the contribution of the isovector mean field can be absorbed in the usual vector mean field and its explicit contribution is hidden.
To achieve the goal of a unified \gls{eos} for both astrophysics and \gls{hic}, the introduction of this term is crucial.
\begin{figure}
	\centering
	\includegraphics[scale=\gpscale]{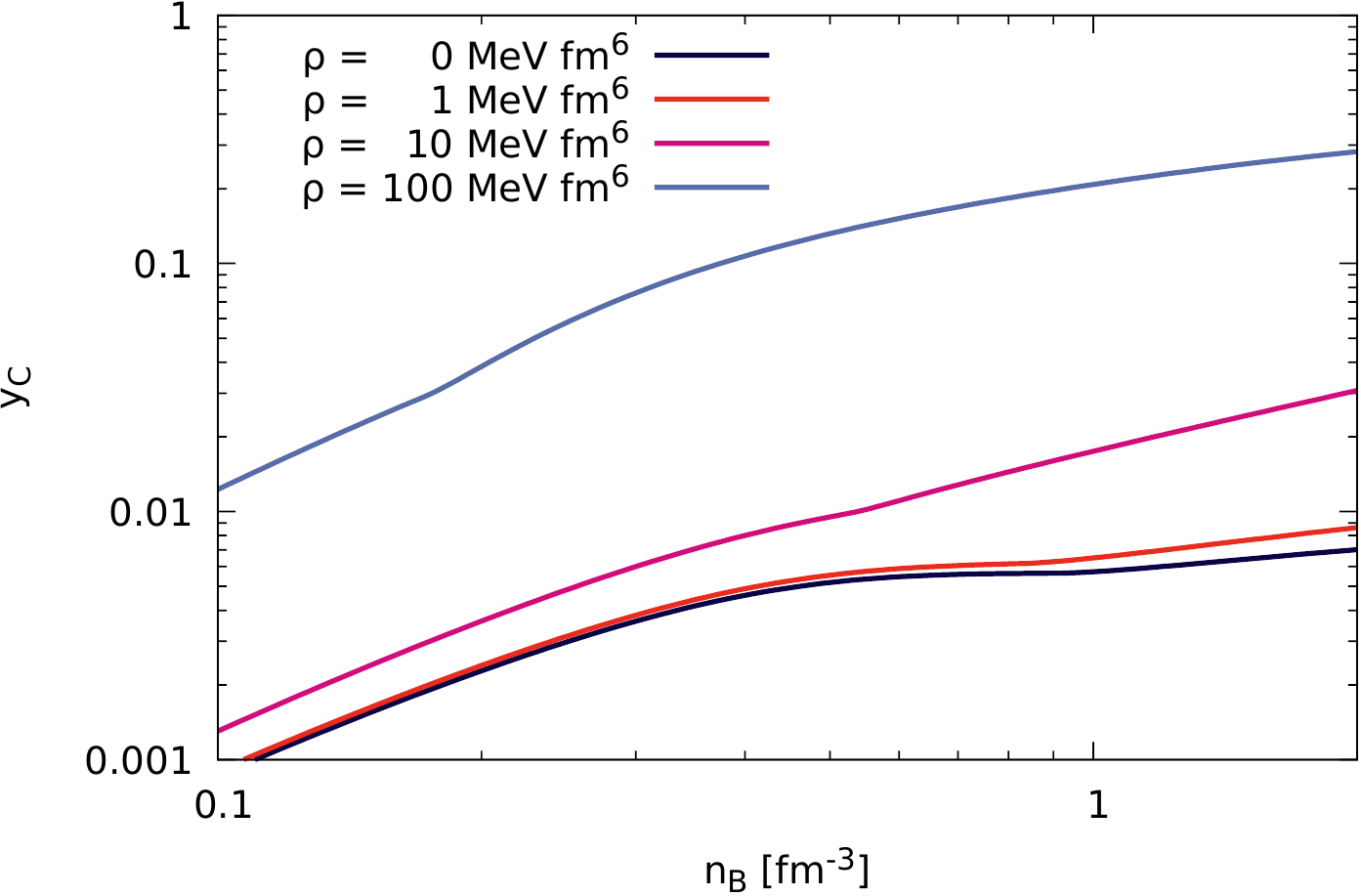}
	\caption{%
	Charge fraction as a function of baryon density at neutron star conditions.
	The different lines correspond to the strength of the vector--isovector parameter $\rho_4$.
	}
	\label{fig:dft:yp}
\end{figure}
Now it is possible to adjust the parameters for symmetric matter applications and constraints independently of the neutron-rich scenarios, e.g., in neutron stars.
Furthermore the shape of the deconfinement phase transition can be adjusted by this contribution, hence is has shown that a model without isovector interaction gives unreasonable onset densities for neutron matter, even below saturation density.
Additionally, the composition of matter in neutron stars, e.g., the baryonic charge fraction
\begin{align}
    \label{eq:model:chargefraction}
	y_\mathrm c &= \frac{1}{n_\mathrm B} \sum_i C_i \nv_i\,,
\end{align}
with $C_i$ being the charge number and $n_\mathrm B$ the total baryon density, strongly depends on this field, as it can be seen in \cref{fig:dft:yp}.
Without any coupling to the vector--isovector mean field, the quark model does not even reach a charge fraction of $1\%$.

\begin{figure}
	\centering
	\includegraphics[scale=\gpscale]{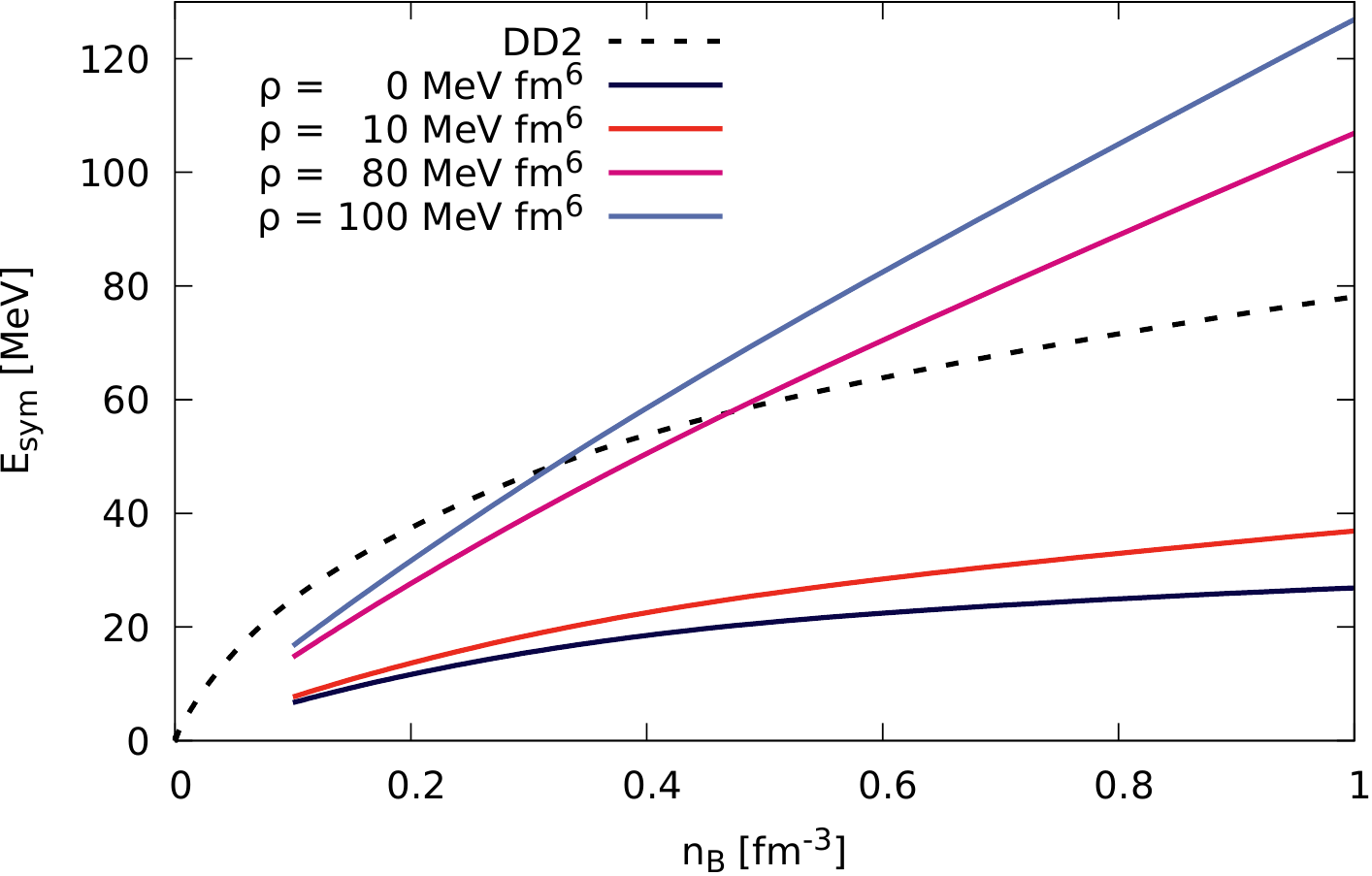}
	\caption{%
	Symmetry energy of quark and hadron matter of hadronic \gls{eos} (DD2) and parametrizations of the quark model with different values of the vector-isovector parameter $\rho_4$ depending on the baryon density.
	}
	\label{fig:dft:esym}
\end{figure}

Still, the question of parametrization is open.
One possibility is to utilise the coupling parameter of a baryonic model and scale by the factor $3$ (``quarks counting rule'').
Another possibility to fit the parameter (which is used, e.g., in \cite{Fischer:2017lag,Bauswein:2018bma}), is to demand an approximately smooth transition of the symmetry energy $E_\mathrm{sym}$ at the deconfinement phase transition, as it can be seen in \cref{fig:dft:esym}.
Without a sufficiently large coupling to the vector--isovector mean field, the symmetry energy would have a significant jump at the quark-hadron phase transition by a factor of up to five.

\subsection{Nuclear Clusters at sub-saturated densities}
\label{sec:model:nuclei}
In astrophysical applications the description of nuclear cluster formation is of special importance.
It does not only change the \gls{eos} itself, but also significantly influences such aspects like neutrino response, for details see, e.g., \citet{Furusawa:2013tta,Fischer:2015sll}.

The current work focuses on the description of high density matter and possible phase transitions to deconfined quark matter.
For the low density problem of nuclear cluster formation, the established model of \citet{Hempel:2009mc} is used.
Here a \gls{nse} is assumed, where any bound state is considered a new species (chemical picture).
The model takes into account experimental values of \citet{Audi:2002rp} and theoretical values of \citet{Moller:1993ed}.
At nuclear saturation density, all clusters should have dissolved and the system should form homogeneous matter of neutrons and protons.
In order to obtain such a Mott transition, the model uses a classic excluded volume approach for the bound states, which suppresses them at higher densities.

\subsection{Leptonic degrees of freedom}
\label{sec:model:leptons}
In the current model, leptons are not taken into account for the phase transition.
In the published tabulations, one version is completely without leptons, because some applications need to add them during the evolution and treat them out of equilibrium.
The other version has electrons added to fulfil electric charge neutrality after the phase transition is constructed.
Muons and neutrinos are not included.

\subsection{Chemical equilibrium and phase transitions}
\label{sec:dft:pt}

In order to obtain a phase transition from hadronic to quark matter, the so-called two-phase approach is applied.
Within this approach, which is commonly used in astrophysics, both phases are derived independently and then merged on a thermodynamic level, applying Gibbs conditions of phase equilibrium.
The fact that hadrons are composite particles made of quarks is not considered in this description.

In the case of chemical equilibrium, the distribution function of all present particles can be expressed by the chemical potentials of their charge numbers as $\mu_i = \sum_j A_{ij} \mu_j$ for each particle $i$, the chemical potential $\mu_j$ of its charges $j$ and the associated charge number $A_{ij}$.
Considered here is baryon number, lepton number and electric charge, resulting in the definition of the particle chemical potentials as
\begin{subequations}
\begin{align}
	\text{Proton: }&&\mu_\mathrm p &= \mu_\mathrm B + \mu_\mathrm C\,,\\
	\text{Neutron: }&&\mu_\mathrm n &= \mu_\mathrm B\,,\\
	\text{Up-Quark: }&&\mu_\mathrm u &= \frac 13\mu_\mathrm B + \frac 23\mu_\mathrm C\,,\\
	\text{Down-Quark: }&&\mu_\mathrm d &= \frac 13\mu_\mathrm B - \frac 13\mu_\mathrm C\,,\\
	\text{Electron: }&&\mu_\mathrm e &= \mu_\mathrm l - \mu_\mathrm C\,,\\
	\text{Neutrino: }&&\mu_{\nu_\mathrm e} &= \mu_\mathrm l\,.
\end{align}
\end{subequations}
Aspects of strangeness are not considered in the current work.
Nevertheless, strangeness is not a conserved charge in astrophysical applications, due to weak equilibrium.
Baryon and charge number are conserved charges.
Lepton number is in astrophysical applications not a conserved charge, and the lepton chemical potential is dictated by charge neutrality.
Furthermore, neutrinos can not always be considered in chemical equilibrium.

Gibbs conditions of phase equilibrium involve thermal equilibrium $T^\mathrm H = T^\mathrm Q$, mechanical equilibrium $p^\mathrm H = p^\mathrm Q$ and chemical equilibrium $\mu_\mathrm{B,C}^\mathrm H = \mu_\mathrm{B,C}^\mathrm Q$ for baryon and charge chemical potential simultaneously.
Since leptons (particularly neutrinos) can not be assumed in equilibrium, they are not taken into account for the phase transition.
At this point, the quark-hadron phase transition is considered to be a phenomenon of strongly interacting particles.

\begin{figure}
	\centering
	\includegraphics[scale=\gpscale]{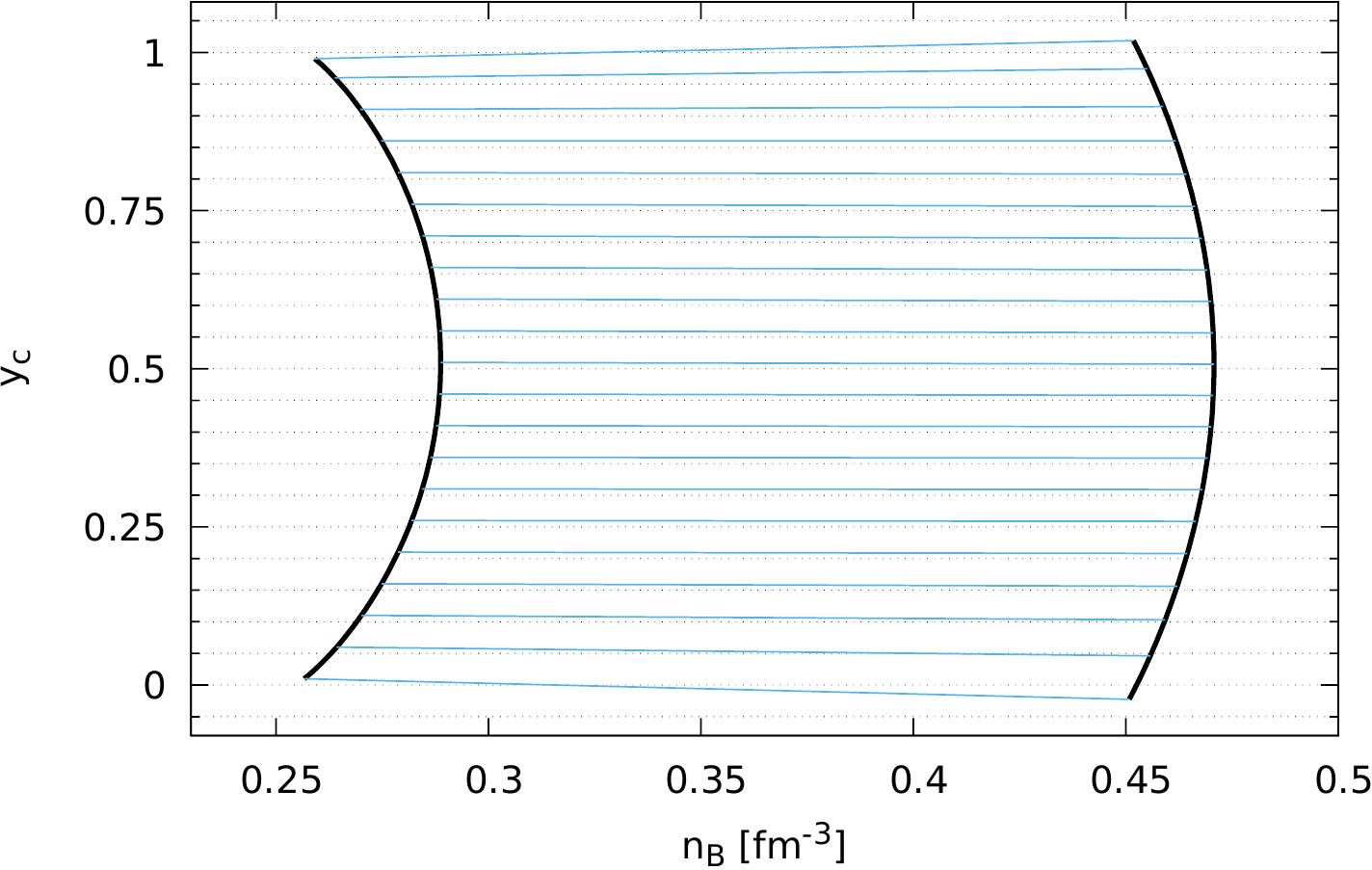}
	\caption{%
    Phase diagram (Charge fraction $y_c$ vs baryon density $n$) of asymmetric matter at zero temperature.
    The boundary of the phase transition is shown as black lines, while the light blue lines connect corresponding quark and hadron points that fulfil the Gibbs conditions.
	}
	\label{fig:model:PDasym}
\end{figure}

In \cref{fig:model:PDasym} the black lines show a resulting phase diagram in the charge fraction over baryon density plane.
For every given charge fraction of the hadronic side, a corresponding \gls{eos} point on the quark side was obtained, which is shown as blue lines.
Generally, these pairs do not have the same charge fraction, because they have equal charge chemical potentials and the dependency of charge chemical potential to charge fraction is system/model dependent.
Note that in two-flavour quark matter the charge fraction can go from $y_\mathrm C = -1$ (pure down quark matter) to $y_\mathrm C = +2$ (pure up quark matter)

The resulting tabulation of the \gls{eos} needs to have isothermal lines of constant charge fraction.
To obtain those, new pairs of points on the phase boundary were selected, which have the same charge fraction but are not in Gibbs-equilibrium.
With those points the mixed phase points were obtained via linear interpolation (in density dimension).

An alternative strategy would be to use the original pairs (as shown in \cref{fig:model:PDasym}) and interpolate until you get the wanted configuration of baryon density and charge fraction.
But despite the additional numerical overhead and errors due to additional interpolations, the effect should be small, due to the small difference in charge fractions in our models.
These small differences are a result of the choice of the parameter $\rho_4$, which controls the symmetry energy of the quark model.
Without this parameter, the effect would be much more drastic.

\section{Results}
\label{sec:results}

\begin{table*}
    \begin{tabular}{lln{3}{0}n{1}{2}n{2}{1}n{1}{1}n{1}{3}n{2}{0}n{1}{3}n{1}{3}n{1}{2}n{1}{2}}
\toprule
Name        &Hadron &{$\sqrt{D_0}$}	&{$\alpha$}	&{$\omega_4$}	&{$\omega_8$}	&{$\omega_8'$}	&{$\rho_4$}	&{$n_1$}	&{$\Delta n$}	&{$M_\mathrm{onset}$}	&{$M_\mathrm{max}$}\\
\midrule
			&   &{[$\mathrm{MeV}$]}	&{[$\mathrm{fm^6}$]}    &{[$\mathrm{MeV\,fm^3}$]}	&{[$\mathrm{MeV\,fm^9}$]}	&{[$\mathrm{fm^6}$]}	&{[$\mathrm{MeV\,fm^3}$]}	&{[$\mathrm{fm^{-3}}$]}	&{[$\mathrm{fm^{-3}}$]} &{[$M_\odot$]} &{[$M_\odot$]}\\
\midrule
\rdf{1}{1}	&DD2F      &265	&0.39	&-4.	&1.6	&0.025	&80	&0.530	&0.109	&1.57	&2.13\\
\rdf{1}{2}	&DD2Fvex   &250	&0.6    &10.	&0.	    &0.	    &80	&0.466	&0.057	&1.37	&2.15\\
\rdf{1}{3}	&DD2F      &240	&0.36	&1.	    &0.5	&0.015	&80	&0.536	&0.125	&1.58	&2.02\\
\rdf{1}{4}	&DD2F      &240	&0.34	&1.	    &0.5	&0.015	&80	&0.579	&0.083	&1.68	&2.02\\
\rdf{1}{5}	&DD2F      &240	&0.38	&1.	    &0.5	&0.015	&80	&0.498	&0.106	&1.48	&2.03\\
\rdf{1}{6}	&DD2F      &240	&0.30	&-3.	&0.8	&0.015	&80	&0.545	&0.120	&1.60	&2.00\\
\rdf{1}{7}	&DD2F      &240	&0.47	&7.	    &0.2	&0.015	&80	&0.562	&0.030	&1.62	&2.11\\
\rdf{1}{8}	&DD2       &240	&0.45	&1.	    &0.5	&0.015	&55	&0.285	&0.255	&0.94	&2.06\\
\rdf{1}{9}	&DD2       &240	&0.63	&10.	&0.	    &0.	    &55	&0.265	&0.189	&0.81	&2.17\\
\bottomrule
\end{tabular}

    \caption{%
    List of parameters for each of the sets presented in the current work.
    The first two columns show the name of the set and the hadronic model, see details in the main text.
    Columns 3-8 present the parameters of the quark model, which are explained in \cref{sec:model:quarks}.
    The last four columns show some representative observables of the parameter sets.
    Here $n_1$ and $\Delta n$ are the onset density and the density jump of the quark-hadron phase transition at neutron star conditions.
    For resulting cold neutron star configurations, the maximal mass $M$ is presented, as well as the lowest mass $M_1$ of neutron stars with a quark core.
    }
    \label{tab:parametersets}
\end{table*}

Main result of this work is the presentation and tabulated publication of a variety of \gls{eos} models, which can be used in astrophysical applications to study the influence of a first-order phase transition at high densities.
Publication of tabulated data will be done on the specialized website \url{http://eos.bastian.science}, which intends to collect all future tabulations for both astrophysical applications and \gls{hic} as well as the CompOSE \footnote{\url{https://compose.obspm.fr}} database, which is a widely used repository in astrophysics, after this work is published.

\subsection{Nomenclature}

Further publications with equation of states are planned, both for astrophysical applications as well as for \gls{hic}, which are based on different hadronic and quark models and can vary in the type of phase-transition.
To form a unified nomenclature of the current and following \gls{eos} the naming scheme \rdf[c]{a}{b} is introduced.
Here $a$ is the index of the set of parametrizations, which mostly have the same formalism and are presented in the same publication, here is always $a=1$.
The index $b$ is the running index inside a published set and $c$ is occurring only in the published file names, in case of technical updates, which do not affect the physics description.

\subsection{Parameters}
In \cref{tab:parametersets} one can find all considered sets of parameters.
The sets \rdf{1}{1} and \rdf{1}{2} are the initial ones, which were developed for supernova simulations by \citet{Fischer:2017lag}.
For the later study of binary neutron star mergers \citep{Bauswein:2018bma} the amount of sets was supplemented by \rdf{1}{3}\dots\rdf{1}{7}, which represent a systematic variation of onset density and latent heat (density jump).
Finally, here two more sets of parameters (\rdf{1}{8} and \rdf{1}{9}) are added to address the demand of a very early onset densities, e.g., to study mergers of hybrid compact stars.

\begin{figure}
	\centering
	\includegraphics[scale=\gpscale]{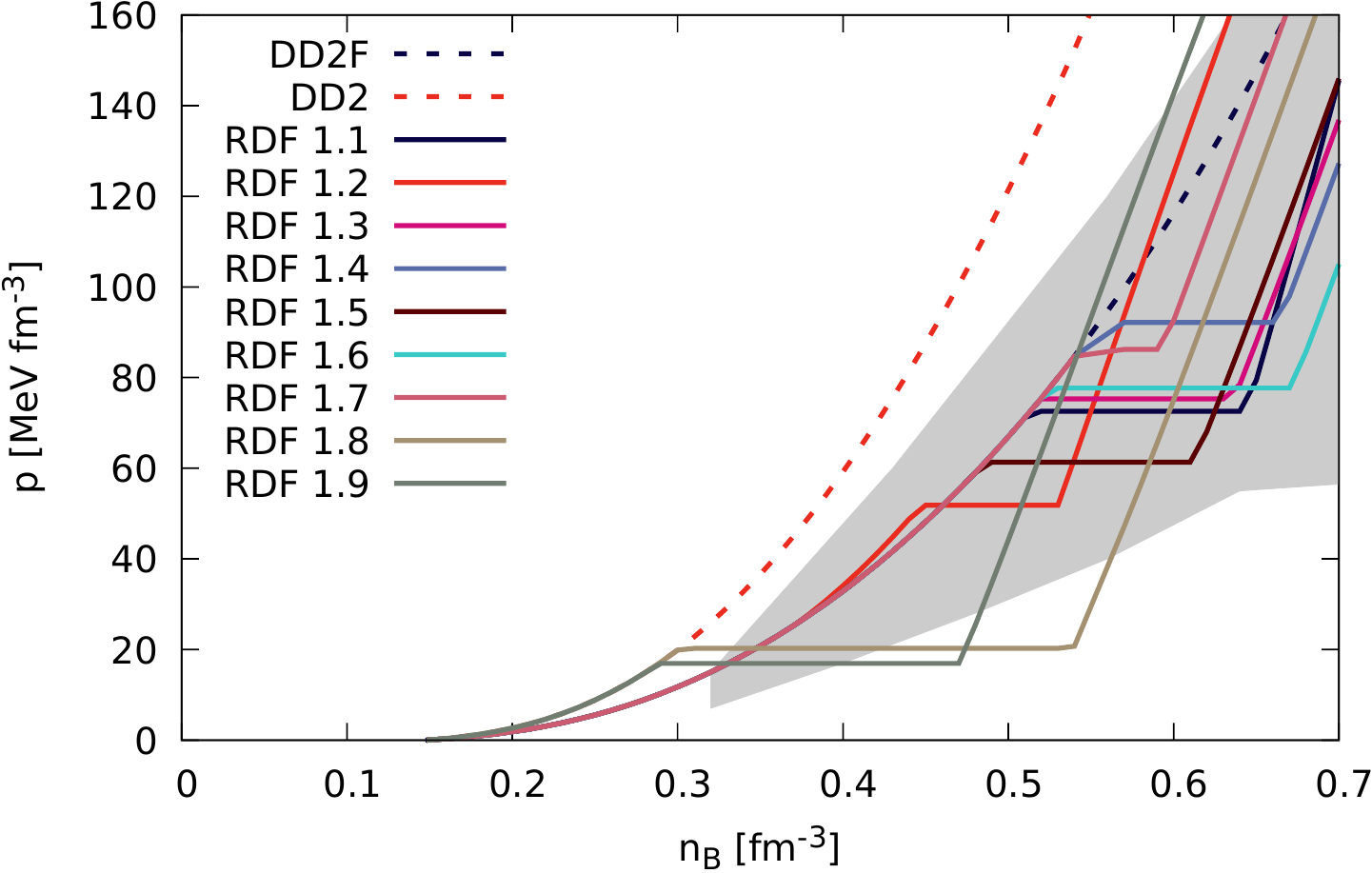}
	\caption{%
	Equation of state shown as pressure $p$ versus baryon density $n_\mathrm B$ for each parameter set.
	The grey shaded area denotes the flow constraint \protect\cite{Danielewicz:2002pu}.
	}
	\label{fig:para:p_n_T0sym}
\end{figure}

\rdf{1}{1}\dots\rdf{1}{7} use the softer DD2F as hadronic model, while \rdf{1}{2} has an excluded volume version of DD2F with its parameters $\alpha^\mathrm{dd2fev} = \np[fm^6]{2.0}$ and $n_0^\mathrm{dd2fev}/n_\mathrm{sat} = 2.5$, but which has only minimal effect on the onset density (visible in \cref{fig:para:p_n_T0sym}).
\rdf{1}{8} and \rdf{1}{9} use the stiffer DD2 to obtain a significantly lower onset density of the phase transition.
The parameters of the quark model are already thoroughly discussed in \cite{Kaltenborn:2017hus}.
Here $\alpha$ varies the onset of the phase transition,  $\omega_4$ is linear and $\omega_8$ is higher order vector coupling, while parameter $\omega_8'$ ensures causality.
The highest-order of vector interaction adjusts the maximal neutron star mass, while the lower order modifies the mixed phase behaviour.
The new parameter $\rho_4$ is the coupling strength to the isovector-vector mean field and therefore varies the behaviour of the symmetry energy, as discussed later.

All sets of parameters are designed to fulfil common constrains.
By the choice of hadronic \gls{eos}  the behaviour until saturation is determined, which fulfils the boundary of \gls{ceft} \citep{Tews:2012fj}, constraints on the symmetry energy and its slope \citep{Lattimer:2012xj,Kolomeitsev:2016sjl} and other properties of saturation density, which are well within the margins of experimental data.
Relevant for densities above saturation density are the maximal mass of neutron stars \cite{Antoniadis:2013pzd,Cromartie:2019kug} and the flow constraint \cite{Danielewicz:2002pu}.
Causality is always preserved, meaning the speed of sound in medium $c_\mathrm{s}^2 = \left(\frac{\partial p}{\partial\varepsilon}\right)_{\{s/n_j\}}$ is always smaller then the speed of light in vacuum.

\begin{figure}
	\centering
	\includegraphics[scale=\gpscale]{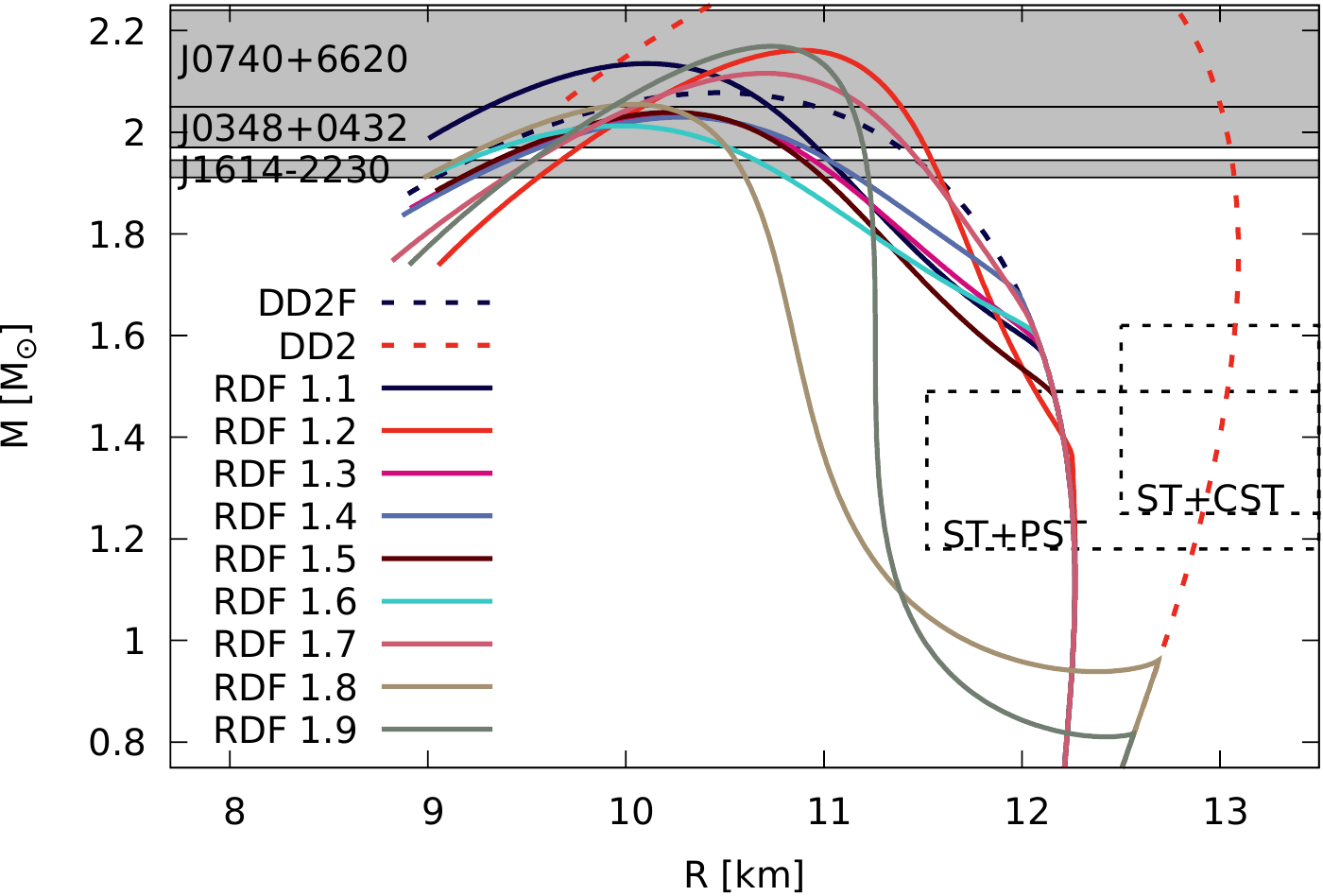}
	\caption{%
	Mass-radius relations $M(R)$ for cold neutron star configurations for each parameter set including the two hadronic reference \gls{eos}.
	Also shown are the experimental constraints of precise high mass measurements and the recent NICER results.
	}
	\label{fig:para:MR_S0}
\end{figure}

An overview of the phase diagrams of all parameter sets can be found in \cref{fig:para:PD_sym}.
The intended variation of low temperature phase transition can be clearly seen in the different panels.
Furthermore, one can see the typical temperature dependant features of two-phase approaches, which are in detail explained in \cref{sec:discussion:Tdependence}.
Note here, that by construction the transition is always of first-order.
This is contradicting the results of \gls{lqcd} calculations \cite{Borsanyi:2013bia,Bazavov:2014pvz}, which predict a smooth crossover transition at vanishing densities.
Therefore, the presented \gls{eos} is not applicable for high energy \gls{hic}, but rather specialized in the investigation of astrophysical phenomena.

In \cref{fig:para:p_n_T0sym} is shown how the \gls{eos} behaves at zero temperature for the isospin symmetric case.
Until the onset density, the \gls{eos} follows exactly the purely hadronic reference curve.
All parametrizations which are based on the \gls{dd2f} \gls{eos} are fulfilling well the flow constraint \citep{Danielewicz:2002pu}.
The hadronic \gls{eos} \gls{dd2} is rather stiff at super-saturated densities and violates the flow constraint.
Anyway, the two parametrizations (\rdf{1}{8} and \rdf{1}{9}), which are based on it, have a phase transition before the applicability of the constraint, resulting in a significant softening.
Because the phase transition is rather big, it is eventually too soft to perfectly lie within the constraint, which is an acceptable discrepancy.

The \cref{fig:para:MR_S0} compares the mass-radius relations for cold neutron stars, together with the highest neutron star masses, which could be measured precisely \citep{Demorest:2010bx,Antoniadis:2013pzd,Cromartie:2019kug}.
Results from radius measurements, done by the \gls{nicer} mission \citep{Raaijmakers:2019qny}, were not taken into account, while creating the parameter sets, but they do not contradict our results within given accuracy.

\begin{figure}
	\centering
	\includegraphics[scale=\gpscale]{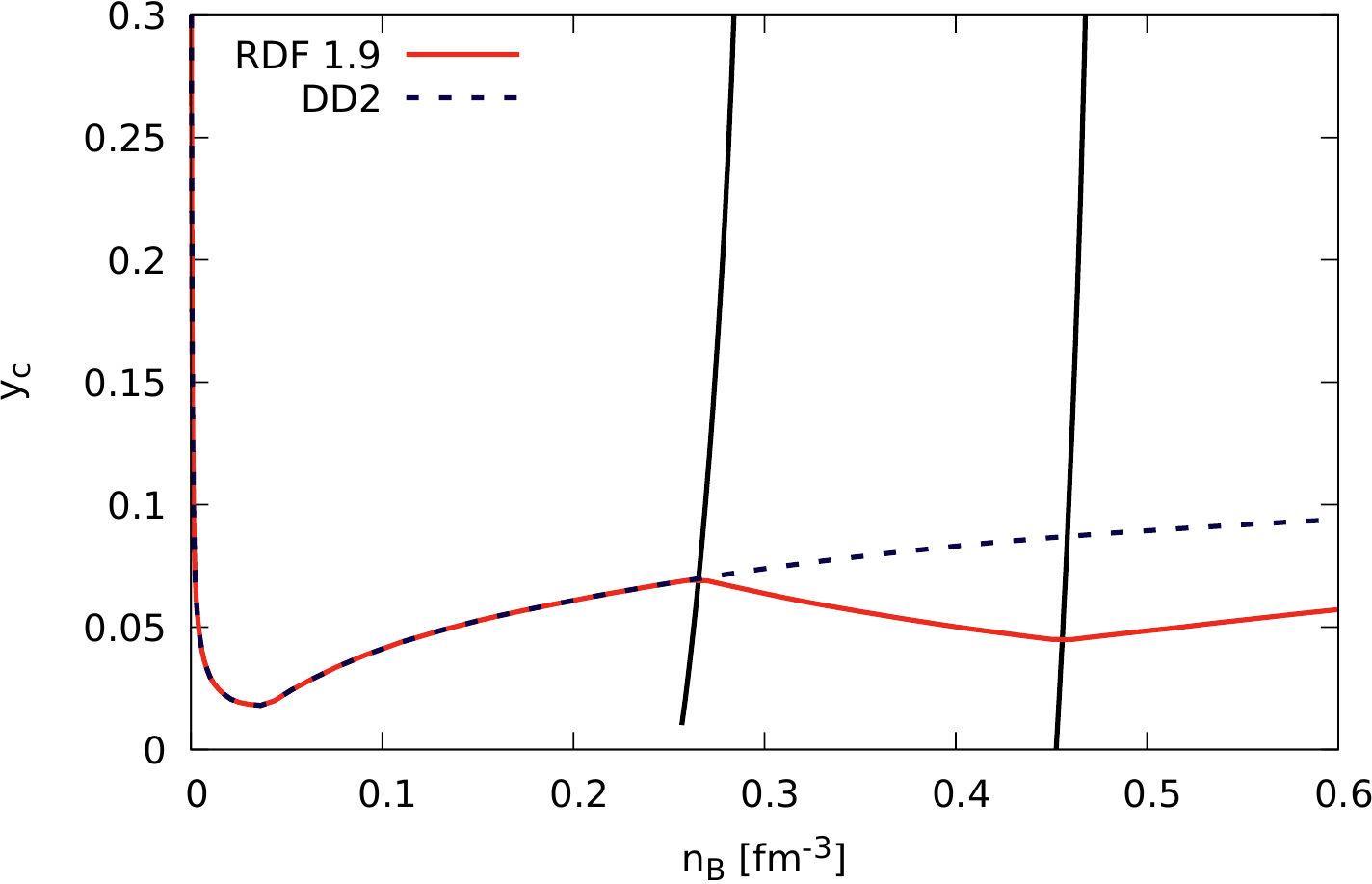}
	\caption{%
	Baryonic charge fraction $y_\mathrm c$ for neutron star conditions (see \cref{eq:model:chargefraction}) for the hadronic reference \gls{eos} \gls{dd2} and the hybrid \gls{eos} \rdf{1}{9}.
	}
	\label{fig:model:pd1}
\end{figure}

The baryonic charge fraction $y_\mathrm C$ for neutron star conditions is shown in \cref{fig:model:pd1}, calculated as an example for RDF 1.9 and its hadronic reference EOS DD2.
As one can see, the two kinks coincide with the phase boundaries, but no discontinuities are observed.
The difference of the charge fraction between DD2 and RDF 1.9 is visible, however it is not as extreme as it would be without vector-isovector couplings (compare with \cref{fig:dft:yp}).

\section{Discussion}
\label{sec:discussion}
Aim of the discussion section is to address particular features which occur in the presented \glspl{eos}.
The discussed parameter set is \rdf{1}{9}, because particular effects are most pronounced.

\subsection{Temperature dependence of the phase transition}
\label{sec:discussion:Tdependence}
\begin{figure}
	\centering
	\includegraphics[scale=\gpscale]{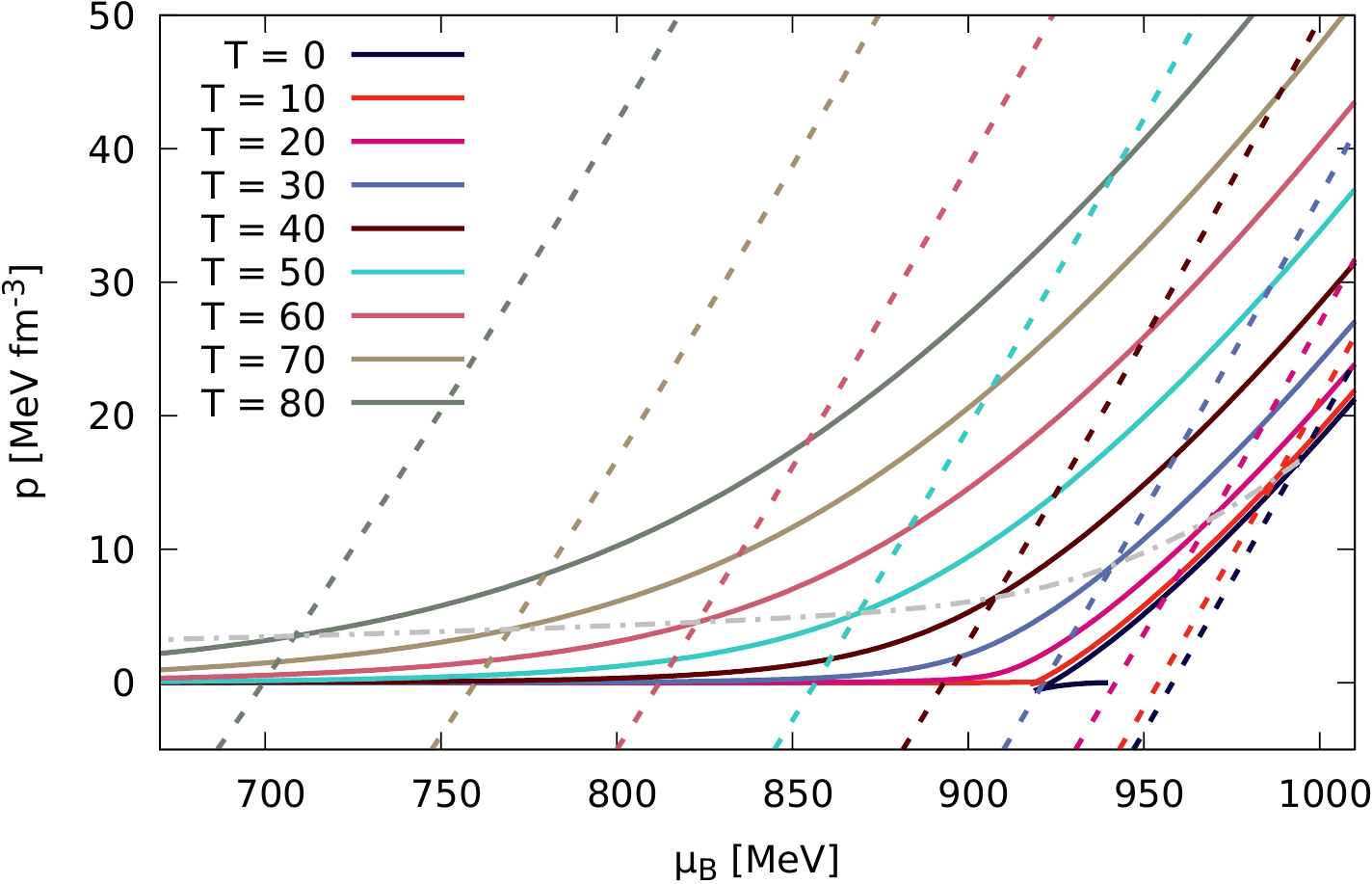}
	\caption{%
	Phase construction for different temperatures in the \rdf{1}{9} set of the model, in the case of symmetric matter.
	The solid lines show the hadronic \gls{eos}, while the corresponding dashed lines are from the quark \gls{eos}.
	The grey dash-dotted line represent the phase transitions, here following the crossings of hadronic and quark \gls{eos} for each value of temperature.
	}
	\label{fig:para:n_mu_Tssym}
\end{figure}

In \cref{fig:para:n_mu_Tssym} it is shown how the phase transition is constructed.
For a given set of temperatures the pressure over baryon chemical potential is shown for both (hadronic and quark) \glspl{eos}.
Here only low and intermediate temperatures are discussed, because at high temperatures the applicability of the two-phase approach ends.
For simplicity only pure symmetric matter is shown, because the charge chemical potential can then be assumed to be zero for both phases.
This simplification is only used in this picture for presentation and not used in the actual data or any other pictures, because the charge chemical potential of the hadronic model is not zero ($\approx \np[MeV]{-1}$), due to the mass difference of neutrons and protons.

The hadronic model features at low temperatures ($T_\mathrm{c} \approx \np[MeV]{14.5}$) the liquid gas phase transition in form of a van-der-Waals wiggle.
For temperatures beyond the critical temperature, the lines are monotonously rising with an increasing gradient.
The quark model always shows a rather steep gradient, starting from negative pressures.
This bag-model-like behaviour originates from the confinement mechanism and is explained in details in \citet{Kaltenborn:2017hus}.
It can also be seen, that the quark model is much more affected by temperature than the hadron model, which can be explained by the lower masses of quarks.

Applying Gibbs conditions and assuming that temperature and charge chemical potential are already equal in both phases, the phase transition is at the crossing point of hadronic and quark line in \cref{fig:para:n_mu_Tssym}.
With rising temperature, one can clearly see, that the model features a monotonous decrease of transition pressure and (baryon) chemical potential.

\begin{figure*}
	\centering
	\includegraphics[scale=\gpscale]{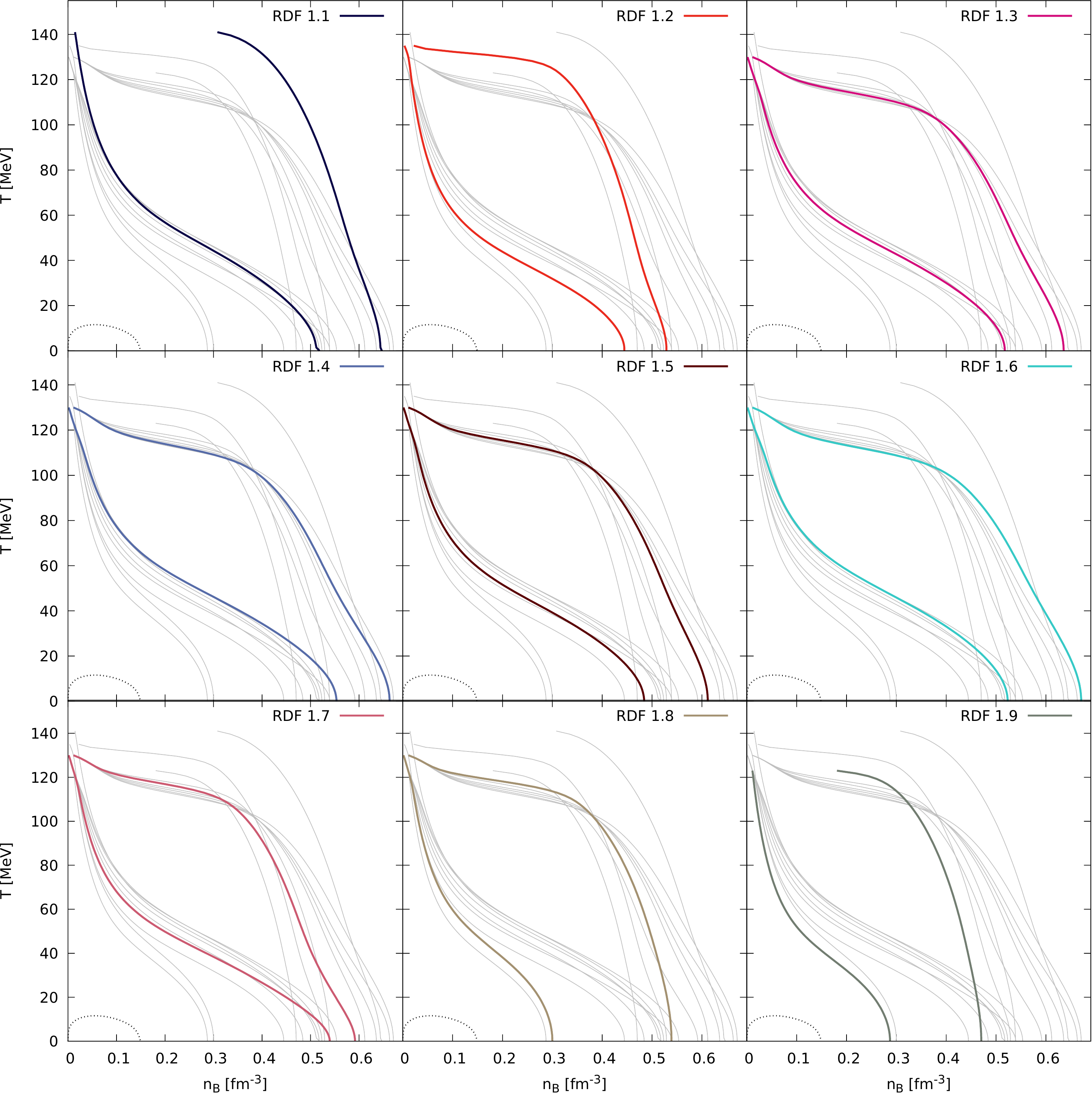}
	\caption{%
	Phase diagrams of symmetric matter for all parameter sets.
	Each panel shows all sets, but highlights a single one.
	Details about the parametrizations can be found in the text.
	The numerical values of the parameters together with characteristic quantities are listed in \cref{tab:parametersets}.
	The lines enclose the mixed phase of the quark-hadron phase transition.
	At lower densities and temperatures one has the pure hadronic phase and at high densities and temperatures the pure quark phase appears.
	The gap at high temperatures, which are most pronounced for \rdf{1}{1} and \rdf{1}{9}, are a numerical artefact due to the sudden drop in density.
	The dotted line at low temperatures and densities is the mixed phase of the liquid-gas phase transition of the hadronic model for comparison.
	}
	\label{fig:para:PD_sym}
\end{figure*}

The baryon density in \cref{fig:para:n_mu_Tssym} can be derived from its gradient $n_\mathrm B = \left(\partial p\big/\partial \mu_\mathrm B\right)_{T,\mu_\mathrm C}$.
Given that the gradient of the quark model, especially around the crossing points, does not have a strong temperature dependence, the transition density from mixed phase to pure quark matter is rather constant w.r.t. temperatures, as it can be seen in the phase diagram in \cref{fig:para:PD_sym}.
The hadron model, on the other hand, has its crossing point at different areas of the \gls{eos}, which feature different gradients.
Therefore, the onset density of the phase transition is highly temperature-dependent.
Note here, that this is a typical feature of two-phase constructions, as discussed in \cref{sec:discussion:2phase}.

\subsection{Leptonic contributions and charge neutrality}
\label{sec:discussion:leptons}

In this work, the \gls{qcd} phase transition is considered a phenomenon of strong interacting particles.
This assumption has several reasons and implications.
One of them is the fact that in dynamical applications, such as supernovae, even electrons can not be considered to be in chemical or even thermal equilibrium with the system.
In such case, the leptons need to be treated separately, e.g. via Boltzmann equation, and the discussion of non-equilibrium effects goes beyond the scope of this manuscript.

Problematic here is the appearance of an additional charge -- the lepton number -- which needs to be taken into account while fulfilling Gibbs conditions, even though in astrophysical applications, the lepton chemical potential is dictated by the assumption of electric charge neutrality of the system.
A most appropriate construction would be done with three chemical potentials, resulting in a four-dimensional (additionally temperature) phase diagram.
For astrophysical applications a three dimensional cut of electric charge neutrality can then be extracted.

If taken into account, at constant chemical potential, the quark \gls{eos} has a higher particle density.
This leads to a stronger contribution (like pressure) by leptons and a resulting lowering of the onset of the phase transition.
Not including leptons during the construction of the phase transition and adding them afterwards always results in a smearing effect, hence all resulting effects are weaker than in a more appropriate treatment and presented results can be seen as conservative.
Eventually the position or even existence of the phase transition is not known and its effects can be studied by systematic variation of also unknown parameters of quark couplings at high density.
The inclusion of leptons would have the need of readjusting these parameters.

\subsection{Description of nuclear clusters}
\label{sec:discussion:nucl}
The current work focuses on the high density part of the equation of state and implications of a possible phase transition to deconfined quark matter.
In order to have clear signatures of these high density features, a well established treatment of the nuclear cluster formation at low densities was chosen.

Direct improvements can be achieved by using updated experimental and theoretical data for nuclear binding energies \citep{Moller:2015fba,Audi:2017asy}.
More sophisticated models are direct quantum-statistical descriptions of light clusters \citep{Ropke:2012qv,Ropke:2014fia,Ropke:2020peo} or generalized density functionals with a depleting binding energy, which significantly change the composition \citep{Typel:2018wmm,Fischer:2020krf}.

\subsection{The two-phase approach}
\label{sec:discussion:2phase}
As was discussed in \cref{sec:dft:pt}, the current work is based on the so-called two-phase approach, where the quark model and the hadron model are completely independent.
Even though it is a commonly used approach, particularly in astrophysics, it has some well-known disadvantages, which significantly limit its usability.

The most obvious problem is that it always gives a first-order transition by construction, which is a contradiction to the \gls{lqcd} results, predicting a smooth crossover at low densities.
In this publication the construction of a first-order transition is the goal and therefore the work is focused on the high-density regime of the \gls{qcd} transition, which is relevant for astrophysics.
Making the model applicable for \gls{hic} would demand a crossover transition at low densities and hence imply a critical endpoint, in case of the existence of a first-order phase transition at high densities.

Another problem, as already discussed by \citet{Bastian:2015avq}, is the fact, that both hadron and quark \gls{eos} are modelled completely separately and their physics does not need to have the same footing.
Effects of one model might occur in the other phase and is by construction suppressed.
An example would be chiral restoration, which is usually modelled for quarks, but might occur already in the hadronic phase \citep{Marczenko:2020jma}.
Furthermore, substructure effects in the hadronic phase like Pauli-blocking are not taken into account, as well as possible remaining bound states or resonances in the quark phase.
The effect of Pauli blocking has been explored in a non-relativistic approximation \citep{Ropke:1986qs,Blaschke:1988zt} and shows similar effects as excluded volume models, e.g., by \citet{Typel:2016srf}.

A solution to most of those problems can be achieved by applying a unified approach, where quarks and hadrons are introduced on the same level, preferably where hadrons are described as bound states of quarks as in the cluster-expansion of \citet{Bastian:2018wfl}.
This approach obtains a \gls{cep}, depending on the parametrization of the interactions \citep{Bastian:2018mmc}.
Substructure effects can be included by the appropriate exchange diagrams.
Finally, a sophisticated formulation of cluster mean field would allow to formulate quark and hadron interactions consistently, which makes it possible to adjust quark interaction parameters to hadronic constraints and have effects like chiral restoration already in the hadronic phase.

A more pragmatic approach is presented by \citet{Typel:2017vif}, where a density-dependent excluded volume formalism is used to achieve a phase transition on a thermodynamic level.
Due to the temperature dependence of this model, it features a critical endpoint at finite temperature.
Since it does not actually describe quarks, this model is limited to applications, which are only sensitive to thermodynamic quantities and not to microscopic quantities.

\section{Summary}
\label{sec:summary}

A microscopical quark-hadron model with an effective interaction potential on the level of the mean-field approximation is presented.
The descriptions of quark matter and hadron matter are done separately and the resulting \glspl{eos} are merged via a thermodynamically consistent two-phase construction.
Dependencies on temperature and baryonic charge fraction are included naturally by use of fermion distribution functions.

Alternative attempts to study the effect of first-order quark-hadron phase transitions in hot astrophysics are done by \citet{Roark:2018uls}, using a chiral mean-field model to describe quark matter and applying a two-phase approach to obtain a first-order phase transition with different incorporation of leptonic degrees of freedom.
The work of \citet{Sagert:2008ka} also applies a two-phase approach for the phase transition, but uses a thermodynamic bag model to describe quark matter, which can not describe neutrons stars of two solar masses and should therefore be considered outdated.
The alternative scenario of a crossover at low temperatures is explored for hyperonic models by \citet{Marques:2017zju} or for quark-hadron \gls{eos} in, e.g., \citet{Baym:2019iky}.

\begin{acknowledgments}
I thank Elizaveta Nazarova, Tobias Fischer, Stefan Typel and David Blaschke for helpful discussions while preparing this manuscript.
This work was funded by the Polish National Science Center (NCN) under grant no. 2019/32/C/ST2/00556.
\end{acknowledgments}

\bibliographystyle{apsrev4-2}
\bibliography{niels_oldstyle}

\begin{thebibliography}{55}%
\makeatletter
\providecommand \@ifxundefined [1]{%
 \@ifx{#1\undefined}
}%
\providecommand \@ifnum [1]{%
 \ifnum #1\expandafter \@firstoftwo
 \else \expandafter \@secondoftwo
 \fi
}%
\providecommand \@ifx [1]{%
 \ifx #1\expandafter \@firstoftwo
 \else \expandafter \@secondoftwo
 \fi
}%
\providecommand \natexlab [1]{#1}%
\providecommand \enquote  [1]{``#1''}%
\providecommand \bibnamefont  [1]{#1}%
\providecommand \bibfnamefont [1]{#1}%
\providecommand \citenamefont [1]{#1}%
\providecommand \href@noop [0]{\@secondoftwo}%
\providecommand \href [0]{\begingroup \@sanitize@url \@href}%
\providecommand \@href[1]{\@@startlink{#1}\@@href}%
\providecommand \@@href[1]{\endgroup#1\@@endlink}%
\providecommand \@sanitize@url [0]{\catcode `\\12\catcode `\$12\catcode
  `\&12\catcode `\#12\catcode `\^12\catcode `\_12\catcode `\%12\relax}%
\providecommand \@@startlink[1]{}%
\providecommand \@@endlink[0]{}%
\providecommand \url  [0]{\begingroup\@sanitize@url \@url }%
\providecommand \@url [1]{\endgroup\@href {#1}{\urlprefix }}%
\providecommand \urlprefix  [0]{URL }%
\providecommand \Eprint [0]{\href }%
\providecommand \doibase [0]{https://doi.org/}%
\providecommand \selectlanguage [0]{\@gobble}%
\providecommand \bibinfo  [0]{\@secondoftwo}%
\providecommand \bibfield  [0]{\@secondoftwo}%
\providecommand \translation [1]{[#1]}%
\providecommand \BibitemOpen [0]{}%
\providecommand \bibitemStop [0]{}%
\providecommand \bibitemNoStop [0]{.\EOS\space}%
\providecommand \EOS [0]{\spacefactor3000\relax}%
\providecommand \BibitemShut  [1]{\csname bibitem#1\endcsname}%
\let\auto@bib@innerbib\@empty
\bibitem [{\citenamefont {Fischer}\ \emph {et~al.}(2017)\citenamefont
  {Fischer}, \citenamefont {Bastian}, \citenamefont {Blaschke}, \citenamefont
  {Cierniak}, \citenamefont {Hempel}, \citenamefont {Klähn}, \citenamefont
  {Martínez-Pinedo}, \citenamefont {Newton}, \citenamefont {Röpke},\ and\
  \citenamefont {Typel}}]{Fischer:2017zcr}%
  \BibitemOpen
  \bibfield  {author} {\bibinfo {author} {\bibfnamefont {T.}~\bibnamefont
  {Fischer}}, \bibinfo {author} {\bibfnamefont {N.-U.}\ \bibnamefont
  {Bastian}}, \bibinfo {author} {\bibfnamefont {D.}~\bibnamefont {Blaschke}},
  \bibinfo {author} {\bibfnamefont {M.}~\bibnamefont {Cierniak}}, \bibinfo
  {author} {\bibfnamefont {M.}~\bibnamefont {Hempel}}, \bibinfo {author}
  {\bibfnamefont {T.}~\bibnamefont {Klähn}}, \bibinfo {author} {\bibfnamefont
  {G.}~\bibnamefont {Martínez-Pinedo}}, \bibinfo {author} {\bibfnamefont
  {W.~G.}\ \bibnamefont {Newton}}, \bibinfo {author} {\bibfnamefont
  {G.}~\bibnamefont {Röpke}},\ and\ \bibinfo {author} {\bibfnamefont
  {S.}~\bibnamefont {Typel}},\ }\href {https://doi.org/10.1017/pasa.2017.63}
  {\bibfield  {journal} {\bibinfo  {journal} {Publ.Astron.Soc.Austral.}\
  }\textbf {\bibinfo {volume} {34}},\ \bibinfo {pages} {e067} (\bibinfo {year}
  {2017})}\BibitemShut {NoStop}%
\bibitem [{\citenamefont {Oertel}\ \emph {et~al.}(2017)\citenamefont {Oertel},
  \citenamefont {Hempel}, \citenamefont {Klähn},\ and\ \citenamefont
  {Typel}}]{Oertel:2016bki}%
  \BibitemOpen
  \bibfield  {author} {\bibinfo {author} {\bibfnamefont {M.}~\bibnamefont
  {Oertel}}, \bibinfo {author} {\bibfnamefont {M.}~\bibnamefont {Hempel}},
  \bibinfo {author} {\bibfnamefont {T.}~\bibnamefont {Klähn}},\ and\ \bibinfo
  {author} {\bibfnamefont {S.}~\bibnamefont {Typel}},\ }\href
  {https://doi.org/10.1103/RevModPhys.89.015007} {\bibfield  {journal}
  {\bibinfo  {journal} {Reviews of Modern Physics}\ }\textbf {\bibinfo {volume}
  {89}},\ \bibinfo {pages} {015007} (\bibinfo {year} {2017})},\ \bibinfo {note}
  {arXiv: 1610.03361}\BibitemShut {NoStop}%
\bibitem [{\citenamefont {Bazavov}\ \emph {et~al.}(2019)\citenamefont
  {Bazavov}, \citenamefont {Ding}, \citenamefont {Hegde}, \citenamefont
  {Kaczmarek}, \citenamefont {Karsch}, \citenamefont {Karthik}, \citenamefont
  {Laermann}, \citenamefont {Lahiri}, \citenamefont {Larsen}, \citenamefont
  {Li}, \citenamefont {Mukherjee}, \citenamefont {Ohno}, \citenamefont
  {Petreczky}, \citenamefont {Sandmeyer}, \citenamefont {Schmidt},
  \citenamefont {Sharma},\ and\ \citenamefont
  {Steinbrecher}}]{Bazavov:2018mes}%
  \BibitemOpen
  \bibfield  {author} {\bibinfo {author} {\bibfnamefont {A.}~\bibnamefont
  {Bazavov}}, \bibinfo {author} {\bibfnamefont {H.-T.}\ \bibnamefont {Ding}},
  \bibinfo {author} {\bibfnamefont {P.}~\bibnamefont {Hegde}}, \bibinfo
  {author} {\bibfnamefont {O.}~\bibnamefont {Kaczmarek}}, \bibinfo {author}
  {\bibfnamefont {F.}~\bibnamefont {Karsch}}, \bibinfo {author} {\bibfnamefont
  {N.}~\bibnamefont {Karthik}}, \bibinfo {author} {\bibfnamefont
  {E.}~\bibnamefont {Laermann}}, \bibinfo {author} {\bibfnamefont
  {A.}~\bibnamefont {Lahiri}}, \bibinfo {author} {\bibfnamefont
  {R.}~\bibnamefont {Larsen}}, \bibinfo {author} {\bibfnamefont {S.-T.}\
  \bibnamefont {Li}}, \bibinfo {author} {\bibfnamefont {S.}~\bibnamefont
  {Mukherjee}}, \bibinfo {author} {\bibfnamefont {H.}~\bibnamefont {Ohno}},
  \bibinfo {author} {\bibfnamefont {P.}~\bibnamefont {Petreczky}}, \bibinfo
  {author} {\bibfnamefont {H.}~\bibnamefont {Sandmeyer}}, \bibinfo {author}
  {\bibfnamefont {C.}~\bibnamefont {Schmidt}}, \bibinfo {author} {\bibfnamefont
  {S.}~\bibnamefont {Sharma}},\ and\ \bibinfo {author} {\bibfnamefont
  {P.}~\bibnamefont {Steinbrecher}},\ }\href
  {https://doi.org/10.1016/j.physletb.2019.05.013} {\bibfield  {journal}
  {\bibinfo  {journal} {Physics Letters B}\ }\textbf {\bibinfo {volume}
  {795}},\ \bibinfo {pages} {15} (\bibinfo {year} {2019})},\ \bibinfo {note}
  {arXiv: 1812.08235}\BibitemShut {NoStop}%
\bibitem [{\citenamefont {Kraemmer}\ and\ \citenamefont
  {Rebhan}(2004)}]{Kraemmer:2003gd}%
  \BibitemOpen
  \bibfield  {author} {\bibinfo {author} {\bibfnamefont {U.}~\bibnamefont
  {Kraemmer}}\ and\ \bibinfo {author} {\bibfnamefont {A.}~\bibnamefont
  {Rebhan}},\ }\href {https://doi.org/10.1088/0034-4885/67/3/R05} {\bibfield
  {journal} {\bibinfo  {journal} {Reports on Progress in Physics}\ }\textbf
  {\bibinfo {volume} {67}},\ \bibinfo {pages} {351} (\bibinfo {year} {2004})},\
  \bibinfo {note} {arXiv: hep-ph/0310337}\BibitemShut {NoStop}%
\bibitem [{\citenamefont {Kurkela}\ \emph {et~al.}(2014)\citenamefont
  {Kurkela}, \citenamefont {Fraga}, \citenamefont {Schaffner-Bielich},\ and\
  \citenamefont {Vuorinen}}]{Kurkela:2014vha}%
  \BibitemOpen
  \bibfield  {author} {\bibinfo {author} {\bibfnamefont {A.}~\bibnamefont
  {Kurkela}}, \bibinfo {author} {\bibfnamefont {E.~S.}\ \bibnamefont {Fraga}},
  \bibinfo {author} {\bibfnamefont {J.}~\bibnamefont {Schaffner-Bielich}},\
  and\ \bibinfo {author} {\bibfnamefont {A.}~\bibnamefont {Vuorinen}},\ }\href
  {https://doi.org/10.1088/0004-637X/789/2/127} {\bibfield  {journal} {\bibinfo
   {journal} {Astrophys.J.}\ }\textbf {\bibinfo {volume} {789}},\ \bibinfo
  {pages} {127} (\bibinfo {year} {2014})}\BibitemShut {NoStop}%
\bibitem [{\citenamefont {Kaltenborn}\ \emph {et~al.}(2017)\citenamefont
  {Kaltenborn}, \citenamefont {Bastian},\ and\ \citenamefont
  {Blaschke}}]{Kaltenborn:2017hus}%
  \BibitemOpen
  \bibfield  {author} {\bibinfo {author} {\bibfnamefont {M.~A.~R.}\
  \bibnamefont {Kaltenborn}}, \bibinfo {author} {\bibfnamefont {N.-U.~F.}\
  \bibnamefont {Bastian}},\ and\ \bibinfo {author} {\bibfnamefont {D.~B.}\
  \bibnamefont {Blaschke}},\ }\bibfield  {journal} {\bibinfo  {journal}
  {Physical Review D}\ }\textbf {\bibinfo {volume} {96}},\ \href
  {https://doi.org/10.1103/PhysRevD.96.056024} {10.1103/PhysRevD.96.056024}
  (\bibinfo {year} {2017})\BibitemShut {NoStop}%
\bibitem [{\citenamefont {Bauswein}\ \emph {et~al.}(2019)\citenamefont
  {Bauswein}, \citenamefont {Bastian}, \citenamefont {Blaschke}, \citenamefont
  {Chatziioannou}, \citenamefont {Clark}, \citenamefont {Fischer},\ and\
  \citenamefont {Oertel}}]{Bauswein:2018bma}%
  \BibitemOpen
  \bibfield  {author} {\bibinfo {author} {\bibfnamefont {A.}~\bibnamefont
  {Bauswein}}, \bibinfo {author} {\bibfnamefont {N.-U.~F.}\ \bibnamefont
  {Bastian}}, \bibinfo {author} {\bibfnamefont {D.~B.}\ \bibnamefont
  {Blaschke}}, \bibinfo {author} {\bibfnamefont {K.}~\bibnamefont
  {Chatziioannou}}, \bibinfo {author} {\bibfnamefont {J.~A.}\ \bibnamefont
  {Clark}}, \bibinfo {author} {\bibfnamefont {T.}~\bibnamefont {Fischer}},\
  and\ \bibinfo {author} {\bibfnamefont {M.}~\bibnamefont {Oertel}},\ }\href
  {https://doi.org/10.1103/PhysRevLett.122.061102} {\bibfield  {journal}
  {\bibinfo  {journal} {Physical Review Letters}\ }\textbf {\bibinfo {volume}
  {122}},\ \bibinfo {pages} {061102} (\bibinfo {year} {2019})},\ \bibinfo
  {note} {publisher: American Physical Society}\BibitemShut {NoStop}%
\bibitem [{\citenamefont {Fischer}\ \emph {et~al.}(2018)\citenamefont
  {Fischer}, \citenamefont {Bastian}, \citenamefont {Wu}, \citenamefont
  {Baklanov}, \citenamefont {Sorokina}, \citenamefont {Blinnikov},
  \citenamefont {Typel}, \citenamefont {Klähn},\ and\ \citenamefont
  {Blaschke}}]{Fischer:2017lag}%
  \BibitemOpen
  \bibfield  {author} {\bibinfo {author} {\bibfnamefont {T.}~\bibnamefont
  {Fischer}}, \bibinfo {author} {\bibfnamefont {N.-U.~F.}\ \bibnamefont
  {Bastian}}, \bibinfo {author} {\bibfnamefont {M.-R.}\ \bibnamefont {Wu}},
  \bibinfo {author} {\bibfnamefont {P.}~\bibnamefont {Baklanov}}, \bibinfo
  {author} {\bibfnamefont {E.}~\bibnamefont {Sorokina}}, \bibinfo {author}
  {\bibfnamefont {S.}~\bibnamefont {Blinnikov}}, \bibinfo {author}
  {\bibfnamefont {S.}~\bibnamefont {Typel}}, \bibinfo {author} {\bibfnamefont
  {T.}~\bibnamefont {Klähn}},\ and\ \bibinfo {author} {\bibfnamefont {D.~B.}\
  \bibnamefont {Blaschke}},\ }\href {https://doi.org/10.1038/s41550-018-0583-0}
  {\bibfield  {journal} {\bibinfo  {journal} {Nat.Astron.}\ }\textbf {\bibinfo
  {volume} {2}},\ \bibinfo {pages} {980} (\bibinfo {year} {2018})}\BibitemShut
  {NoStop}%
\bibitem [{\citenamefont {Fischer}\ \emph
  {et~al.}(2020{\natexlab{a}})\citenamefont {Fischer}, \citenamefont {Wu},
  \citenamefont {Wehmeyer}, \citenamefont {Bastian}, \citenamefont
  {Martínez-Pinedo},\ and\ \citenamefont {Thielemann}}]{Fischer:2020xjl}%
  \BibitemOpen
  \bibfield  {author} {\bibinfo {author} {\bibfnamefont {T.}~\bibnamefont
  {Fischer}}, \bibinfo {author} {\bibfnamefont {M.-R.}\ \bibnamefont {Wu}},
  \bibinfo {author} {\bibfnamefont {B.}~\bibnamefont {Wehmeyer}}, \bibinfo
  {author} {\bibfnamefont {N.-U.~F.}\ \bibnamefont {Bastian}}, \bibinfo
  {author} {\bibfnamefont {G.}~\bibnamefont {Martínez-Pinedo}},\ and\ \bibinfo
  {author} {\bibfnamefont {F.-K.}\ \bibnamefont {Thielemann}},\ }\href
  {https://doi.org/10.3847/1538-4357/ab86b0} {\bibfield  {journal} {\bibinfo
  {journal} {The Astrophysical Journal}\ }\textbf {\bibinfo {volume} {894}},\
  \bibinfo {pages} {9} (\bibinfo {year} {2020}{\natexlab{a}})},\ \bibinfo
  {note} {arXiv: 2003.00972}\BibitemShut {NoStop}%
\bibitem [{\citenamefont {Kapusta}(1989)}]{Kapusta:1989tk}%
  \BibitemOpen
  \bibfield  {author} {\bibinfo {author} {\bibfnamefont {J.~I.}\ \bibnamefont
  {Kapusta}},\ }\href@noop {} {\emph {\bibinfo {title} {Finite temperature
  field theory}}},\ \bibinfo {series} {Cambridge monographs on mathematical
  physics}, Vol.\ \bibinfo {volume} {360}\ (\bibinfo  {publisher} {Cambridge
  University Press},\ \bibinfo {address} {Cambridge},\ \bibinfo {year}
  {1989})\BibitemShut {NoStop}%
\bibitem [{\citenamefont {Typel}\ and\ \citenamefont
  {Wolter}(1999)}]{Typel:1999yq}%
  \BibitemOpen
  \bibfield  {author} {\bibinfo {author} {\bibfnamefont {S.}~\bibnamefont
  {Typel}}\ and\ \bibinfo {author} {\bibfnamefont {H.~H.}\ \bibnamefont
  {Wolter}},\ }\href {https://doi.org/10.1016/S0375-9474(99)00310-3} {\bibfield
   {journal} {\bibinfo  {journal} {Nucl.Phys.}\ }\textbf {\bibinfo {volume}
  {A656}},\ \bibinfo {pages} {331} (\bibinfo {year} {1999})}\BibitemShut
  {NoStop}%
\bibitem [{\citenamefont {Typel}\ \emph {et~al.}(2010)\citenamefont {Typel},
  \citenamefont {Ropke}, \citenamefont {Klahn}, \citenamefont {Blaschke},\ and\
  \citenamefont {Wolter}}]{Typel:2009sy}%
  \BibitemOpen
  \bibfield  {author} {\bibinfo {author} {\bibfnamefont {S.}~\bibnamefont
  {Typel}}, \bibinfo {author} {\bibfnamefont {G.}~\bibnamefont {Ropke}},
  \bibinfo {author} {\bibfnamefont {T.}~\bibnamefont {Klahn}}, \bibinfo
  {author} {\bibfnamefont {D.}~\bibnamefont {Blaschke}},\ and\ \bibinfo
  {author} {\bibfnamefont {H.~H.}\ \bibnamefont {Wolter}},\ }\href
  {https://doi.org/10.1103/PhysRevC.81.015803} {\bibfield  {journal} {\bibinfo
  {journal} {Phys.Rev.}\ }\textbf {\bibinfo {volume} {C81}},\ \bibinfo {pages}
  {015803} (\bibinfo {year} {2010})}\BibitemShut {NoStop}%
\bibitem [{\citenamefont {Danielewicz}\ \emph {et~al.}(2002)\citenamefont
  {Danielewicz}, \citenamefont {Lacey},\ and\ \citenamefont
  {Lynch}}]{Danielewicz:2002pu}%
  \BibitemOpen
  \bibfield  {author} {\bibinfo {author} {\bibfnamefont {P.}~\bibnamefont
  {Danielewicz}}, \bibinfo {author} {\bibfnamefont {R.}~\bibnamefont {Lacey}},\
  and\ \bibinfo {author} {\bibfnamefont {W.~G.}\ \bibnamefont {Lynch}},\ }\href
  {https://doi.org/10.1126/science.1078070} {\bibfield  {journal} {\bibinfo
  {journal} {Science}\ }\textbf {\bibinfo {volume} {298}},\ \bibinfo {pages}
  {1592} (\bibinfo {year} {2002})}\BibitemShut {NoStop}%
\bibitem [{\citenamefont {Alvarez-Castillo}\ \emph {et~al.}(2016)\citenamefont
  {Alvarez-Castillo}, \citenamefont {Ayriyan}, \citenamefont {Benic},
  \citenamefont {Blaschke}, \citenamefont {Grigorian},\ and\ \citenamefont
  {Typel}}]{Alvarez-Castillo:2016oln}%
  \BibitemOpen
  \bibfield  {author} {\bibinfo {author} {\bibfnamefont {D.}~\bibnamefont
  {Alvarez-Castillo}}, \bibinfo {author} {\bibfnamefont {A.}~\bibnamefont
  {Ayriyan}}, \bibinfo {author} {\bibfnamefont {S.}~\bibnamefont {Benic}},
  \bibinfo {author} {\bibfnamefont {D.}~\bibnamefont {Blaschke}}, \bibinfo
  {author} {\bibfnamefont {H.}~\bibnamefont {Grigorian}},\ and\ \bibinfo
  {author} {\bibfnamefont {S.}~\bibnamefont {Typel}},\ }\href
  {https://doi.org/10.1140/epja/i2016-16069-2} {\bibfield  {journal} {\bibinfo
  {journal} {The European Physical Journal A}\ }\textbf {\bibinfo {volume}
  {52}},\ \bibinfo {pages} {69} (\bibinfo {year} {2016})}\BibitemShut {NoStop}%
\bibitem [{\citenamefont {Typel}\ and\ \citenamefont
  {Alvear~Terrero}(2020)}]{Typel:2020ozc}%
  \BibitemOpen
  \bibfield  {author} {\bibinfo {author} {\bibfnamefont {S.}~\bibnamefont
  {Typel}}\ and\ \bibinfo {author} {\bibfnamefont {D.}~\bibnamefont
  {Alvear~Terrero}},\ }\href {https://doi.org/10.1140/epja/s10050-020-00172-2}
  {\bibfield  {journal} {\bibinfo  {journal} {The European Physical Journal A}\
  }\textbf {\bibinfo {volume} {56}},\ \bibinfo {pages} {160} (\bibinfo {year}
  {2020})},\ \bibinfo {note} {arXiv: 2003.02085}\BibitemShut {NoStop}%
\bibitem [{\citenamefont {Glozman}(2009)}]{Glozman:2008fk}%
  \BibitemOpen
  \bibfield  {author} {\bibinfo {author} {\bibfnamefont {L.~Y.}\ \bibnamefont
  {Glozman}},\ }\href {https://doi.org/10.1103/PhysRevD.79.037504} {\bibfield
  {journal} {\bibinfo  {journal} {Physical Review D}\ }\textbf {\bibinfo
  {volume} {79}},\ \bibinfo {pages} {037504} (\bibinfo {year}
  {2009})}\BibitemShut {NoStop}%
\bibitem [{\citenamefont {Li}\ \emph {et~al.}(2015)\citenamefont {Li},
  \citenamefont {Zuo},\ and\ \citenamefont {Peng}}]{Li:2015ida}%
  \BibitemOpen
  \bibfield  {author} {\bibinfo {author} {\bibfnamefont {A.}~\bibnamefont
  {Li}}, \bibinfo {author} {\bibfnamefont {W.}~\bibnamefont {Zuo}},\ and\
  \bibinfo {author} {\bibfnamefont {G.~X.}\ \bibnamefont {Peng}},\ }\href
  {https://doi.org/10.1103/PhysRevC.91.035803} {\bibfield  {journal} {\bibinfo
  {journal} {Phys.Rev.}\ }\textbf {\bibinfo {volume} {C91}},\ \bibinfo {pages}
  {035803} (\bibinfo {year} {2015})}\BibitemShut {NoStop}%
\bibitem [{\citenamefont {Ropke}\ \emph {et~al.}(1986)\citenamefont {Ropke},
  \citenamefont {Blaschke},\ and\ \citenamefont {Schulz}}]{Ropke:1986qs}%
  \BibitemOpen
  \bibfield  {author} {\bibinfo {author} {\bibfnamefont {G.}~\bibnamefont
  {Ropke}}, \bibinfo {author} {\bibfnamefont {D.}~\bibnamefont {Blaschke}},\
  and\ \bibinfo {author} {\bibfnamefont {H.}~\bibnamefont {Schulz}},\ }\href
  {https://doi.org/10.1103/PhysRevD.34.3499} {\bibfield  {journal} {\bibinfo
  {journal} {Phys.Rev.}\ }\textbf {\bibinfo {volume} {D34}},\ \bibinfo {pages}
  {3499} (\bibinfo {year} {1986})}\BibitemShut {NoStop}%
\bibitem [{\citenamefont {Blaschke}\ \emph {et~al.}(1990)\citenamefont
  {Blaschke}, \citenamefont {Tovmasian},\ and\ \citenamefont
  {Kampfer}}]{Blaschke:1989nn}%
  \BibitemOpen
  \bibfield  {author} {\bibinfo {author} {\bibfnamefont {D.}~\bibnamefont
  {Blaschke}}, \bibinfo {author} {\bibfnamefont {T.}~\bibnamefont
  {Tovmasian}},\ and\ \bibinfo {author} {\bibfnamefont {B.}~\bibnamefont
  {Kampfer}},\ }\href@noop {} {\bibfield  {journal} {\bibinfo  {journal}
  {Soviet Journal of Nuclear Physics}\ }\textbf {\bibinfo {volume} {52}},\
  \bibinfo {pages} {675} (\bibinfo {year} {1990})}\ \bibinfo {note} {Preprint: JINR-E2-89-651}\BibitemShut {NoStop}%
\bibitem [{\citenamefont {Serot}\ and\ \citenamefont
  {Walecka}(1997)}]{Serot:1997xg}%
  \BibitemOpen
  \bibfield  {author} {\bibinfo {author} {\bibfnamefont {B.~D.}\ \bibnamefont
  {Serot}}\ and\ \bibinfo {author} {\bibfnamefont {J.~D.}\ \bibnamefont
  {Walecka}},\ }\href {https://doi.org/10.1142/S0218301397000299} {\bibfield
  {journal} {\bibinfo  {journal} {International Journal of Modern Physics E:
  Nuclear Physics}\ }\textbf {\bibinfo {volume} {6}},\ \bibinfo {pages} {515}
  (\bibinfo {year} {1997})},\ \bibinfo {note} {arXiv:
  nucl-th/9701058}\BibitemShut {NoStop}%
\bibitem [{\citenamefont {Benic}(2014)}]{Benic:2014iaa}%
  \BibitemOpen
  \bibfield  {author} {\bibinfo {author} {\bibfnamefont {S.}~\bibnamefont
  {Benic}},\ }\href {https://doi.org/10.1140/epja/i2014-14111-1} {\bibfield
  {journal} {\bibinfo  {journal} {European Physical Journal A: Hadrons and
  Nuclei}\ }\textbf {\bibinfo {volume} {50}},\ \bibinfo {pages} {111} (\bibinfo
  {year} {2014})},\ \bibinfo {note} {arXiv: 1401.5380 [nucl-th]}\BibitemShut
  {NoStop}%
\bibitem [{\citenamefont {Demorest}\ \emph {et~al.}(2010)\citenamefont
  {Demorest}, \citenamefont {Pennucci}, \citenamefont {Ransom}, \citenamefont
  {Roberts},\ and\ \citenamefont {Hessels}}]{Demorest:2010bx}%
  \BibitemOpen
  \bibfield  {author} {\bibinfo {author} {\bibfnamefont {P.}~\bibnamefont
  {Demorest}}, \bibinfo {author} {\bibfnamefont {T.}~\bibnamefont {Pennucci}},
  \bibinfo {author} {\bibfnamefont {S.}~\bibnamefont {Ransom}}, \bibinfo
  {author} {\bibfnamefont {M.}~\bibnamefont {Roberts}},\ and\ \bibinfo {author}
  {\bibfnamefont {J.}~\bibnamefont {Hessels}},\ }\href
  {https://doi.org/10.1038/nature09466} {\bibfield  {journal} {\bibinfo
  {journal} {Nature}\ }\textbf {\bibinfo {volume} {467}},\ \bibinfo {pages}
  {1081} (\bibinfo {year} {2010})}\BibitemShut {NoStop}%
\bibitem [{\citenamefont {Antoniadis}\ \emph {et~al.}(2013)\citenamefont
  {Antoniadis}, \citenamefont {Freire}, \citenamefont {Wex}, \citenamefont
  {Tauris}, \citenamefont {Lynch}, \citenamefont {van Kerkwijk}, \citenamefont
  {Kramer}, \citenamefont {Bassa}, \citenamefont {Dhillon}, \citenamefont
  {Driebe}, \citenamefont {Hessels}, \citenamefont {Kaspi}, \citenamefont
  {Kondratiev}, \citenamefont {Langer}, \citenamefont {Marsh}, \citenamefont
  {McLaughlin}, \citenamefont {Pennucci}, \citenamefont {Ransom}, \citenamefont
  {Stairs}, \citenamefont {van Leeuwen}, \citenamefont {Verbiest},\ and\
  \citenamefont {Whelan}}]{Antoniadis:2013pzd}%
  \BibitemOpen
  \bibfield  {author} {\bibinfo {author} {\bibfnamefont {J.}~\bibnamefont
  {Antoniadis}}, \bibinfo {author} {\bibfnamefont {P.~C.~C.}\ \bibnamefont
  {Freire}}, \bibinfo {author} {\bibfnamefont {N.}~\bibnamefont {Wex}},
  \bibinfo {author} {\bibfnamefont {T.~M.}\ \bibnamefont {Tauris}}, \bibinfo
  {author} {\bibfnamefont {R.~S.}\ \bibnamefont {Lynch}}, \bibinfo {author}
  {\bibfnamefont {M.~H.}\ \bibnamefont {van Kerkwijk}}, \bibinfo {author}
  {\bibfnamefont {M.}~\bibnamefont {Kramer}}, \bibinfo {author} {\bibfnamefont
  {C.}~\bibnamefont {Bassa}}, \bibinfo {author} {\bibfnamefont {V.~S.}\
  \bibnamefont {Dhillon}}, \bibinfo {author} {\bibfnamefont {T.}~\bibnamefont
  {Driebe}}, \bibinfo {author} {\bibfnamefont {J.~W.~T.}\ \bibnamefont
  {Hessels}}, \bibinfo {author} {\bibfnamefont {V.~M.}\ \bibnamefont {Kaspi}},
  \bibinfo {author} {\bibfnamefont {V.~I.}\ \bibnamefont {Kondratiev}},
  \bibinfo {author} {\bibfnamefont {N.}~\bibnamefont {Langer}}, \bibinfo
  {author} {\bibfnamefont {T.~R.}\ \bibnamefont {Marsh}}, \bibinfo {author}
  {\bibfnamefont {M.~A.}\ \bibnamefont {McLaughlin}}, \bibinfo {author}
  {\bibfnamefont {T.~T.}\ \bibnamefont {Pennucci}}, \bibinfo {author}
  {\bibfnamefont {S.~M.}\ \bibnamefont {Ransom}}, \bibinfo {author}
  {\bibfnamefont {I.~H.}\ \bibnamefont {Stairs}}, \bibinfo {author}
  {\bibfnamefont {J.}~\bibnamefont {van Leeuwen}}, \bibinfo {author}
  {\bibfnamefont {J.~P.~W.}\ \bibnamefont {Verbiest}},\ and\ \bibinfo {author}
  {\bibfnamefont {D.~G.}\ \bibnamefont {Whelan}},\ }\href
  {https://doi.org/10.1126/science.1233232} {\bibfield  {journal} {\bibinfo
  {journal} {Science}\ }\textbf {\bibinfo {volume} {340}},\ \bibinfo {pages}
  {1233232} (\bibinfo {year} {2013})}\BibitemShut {NoStop}%
\bibitem [{\citenamefont {Benic}\ \emph {et~al.}(2015)\citenamefont {Benic},
  \citenamefont {Blaschke}, \citenamefont {Alvarez-Castillo}, \citenamefont
  {Fischer},\ and\ \citenamefont {Typel}}]{Benic:2014jia}%
  \BibitemOpen
  \bibfield  {author} {\bibinfo {author} {\bibfnamefont {S.}~\bibnamefont
  {Benic}}, \bibinfo {author} {\bibfnamefont {D.}~\bibnamefont {Blaschke}},
  \bibinfo {author} {\bibfnamefont {D.~E.}\ \bibnamefont {Alvarez-Castillo}},
  \bibinfo {author} {\bibfnamefont {T.}~\bibnamefont {Fischer}},\ and\ \bibinfo
  {author} {\bibfnamefont {S.}~\bibnamefont {Typel}},\ }\href
  {https://doi.org/10.1051/0004-6361/201425318} {\bibfield  {journal} {\bibinfo
   {journal} {Astron.Astrophys.}\ }\textbf {\bibinfo {volume} {577}},\ \bibinfo
  {pages} {A40} (\bibinfo {year} {2015})}\BibitemShut {NoStop}%
\bibitem [{\citenamefont {Furusawa}\ \emph {et~al.}(2013)\citenamefont
  {Furusawa}, \citenamefont {Nagakura}, \citenamefont {Sumiyoshi},\ and\
  \citenamefont {Yamada}}]{Furusawa:2013tta}%
  \BibitemOpen
  \bibfield  {author} {\bibinfo {author} {\bibfnamefont {S.}~\bibnamefont
  {Furusawa}}, \bibinfo {author} {\bibfnamefont {H.}~\bibnamefont {Nagakura}},
  \bibinfo {author} {\bibfnamefont {K.}~\bibnamefont {Sumiyoshi}},\ and\
  \bibinfo {author} {\bibfnamefont {S.}~\bibnamefont {Yamada}},\ }\href
  {https://doi.org/10.1088/0004-637X/774/1/78} {\bibfield  {journal} {\bibinfo
  {journal} {The Astrophysical Journal}\ }\textbf {\bibinfo {volume} {774}},\
  \bibinfo {pages} {78} (\bibinfo {year} {2013})},\ \bibinfo {note} {arXiv:
  1305.1510}\BibitemShut {NoStop}%
\bibitem [{\citenamefont {Fischer}\ \emph {et~al.}(2016)\citenamefont
  {Fischer}, \citenamefont {Martínez-Pinedo}, \citenamefont {Hempel},
  \citenamefont {Huther}, \citenamefont {Röpke}, \citenamefont {Typel},\ and\
  \citenamefont {Lohs}}]{Fischer:2015sll}%
  \BibitemOpen
  \bibfield  {author} {\bibinfo {author} {\bibfnamefont {T.}~\bibnamefont
  {Fischer}}, \bibinfo {author} {\bibfnamefont {G.}~\bibnamefont
  {Martínez-Pinedo}}, \bibinfo {author} {\bibfnamefont {M.}~\bibnamefont
  {Hempel}}, \bibinfo {author} {\bibfnamefont {L.}~\bibnamefont {Huther}},
  \bibinfo {author} {\bibfnamefont {G.}~\bibnamefont {Röpke}}, \bibinfo
  {author} {\bibfnamefont {S.}~\bibnamefont {Typel}},\ and\ \bibinfo {author}
  {\bibfnamefont {A.}~\bibnamefont {Lohs}},\ }\href
  {https://doi.org/10.1051/epjconf/201610906002} {\bibfield  {journal}
  {\bibinfo  {journal} {EPJ Web Conf.}\ }\textbf {\bibinfo {volume} {109}},\
  \bibinfo {pages} {06002} (\bibinfo {year} {2016})}\BibitemShut {NoStop}%
\bibitem [{\citenamefont {Hempel}\ and\ \citenamefont
  {Schaffner-Bielich}(2010)}]{Hempel:2009mc}%
  \BibitemOpen
  \bibfield  {author} {\bibinfo {author} {\bibfnamefont {M.}~\bibnamefont
  {Hempel}}\ and\ \bibinfo {author} {\bibfnamefont {J.}~\bibnamefont
  {Schaffner-Bielich}},\ }\href
  {https://doi.org/10.1016/j.nuclphysa.2010.02.010} {\bibfield  {journal}
  {\bibinfo  {journal} {Nucl.Phys.}\ }\textbf {\bibinfo {volume} {A837}},\
  \bibinfo {pages} {210} (\bibinfo {year} {2010})}\BibitemShut {NoStop}%
\bibitem [{\citenamefont {Audi}\ \emph {et~al.}(2002)\citenamefont {Audi},
  \citenamefont {Wapstra},\ and\ \citenamefont {Thibault}}]{Audi:2002rp}%
  \BibitemOpen
  \bibfield  {author} {\bibinfo {author} {\bibfnamefont {G.}~\bibnamefont
  {Audi}}, \bibinfo {author} {\bibfnamefont {A.}~\bibnamefont {Wapstra}},\ and\
  \bibinfo {author} {\bibfnamefont {C.}~\bibnamefont {Thibault}},\ }\href
  {https://doi.org/10.1016/j.nuclphysa.2003.11.003} {\bibfield  {journal}
  {\bibinfo  {journal} {Nuclear Physics A}\ }\textbf {\bibinfo {volume}
  {729}},\ \bibinfo {pages} {337} (\bibinfo {year} {2002})}\BibitemShut
  {NoStop}%
\bibitem [{\citenamefont {Moller}\ \emph {et~al.}(1995)\citenamefont {Moller},
  \citenamefont {Nix}, \citenamefont {Myers},\ and\ \citenamefont
  {Swiatecki}}]{Moller:1993ed}%
  \BibitemOpen
  \bibfield  {author} {\bibinfo {author} {\bibfnamefont {P.}~\bibnamefont
  {Moller}}, \bibinfo {author} {\bibfnamefont {J.}~\bibnamefont {Nix}},
  \bibinfo {author} {\bibfnamefont {W.}~\bibnamefont {Myers}},\ and\ \bibinfo
  {author} {\bibfnamefont {W.}~\bibnamefont {Swiatecki}},\ }\href
  {https://doi.org/10.1006/adnd.1995.1002} {\bibfield  {journal} {\bibinfo
  {journal} {Atom. Data Nucl. Data Tabl.}\ }\textbf {\bibinfo {volume} {59}},\
  \bibinfo {pages} {185} (\bibinfo {year} {1995})},\ \bibinfo {note} {arXiv:
  nucl-th/9308022}\BibitemShut {NoStop}%
\bibitem [{Note1()}]{Note1}%
  \BibitemOpen
  \bibinfo {note} {\protect \url {https://compose.obspm.fr}}\BibitemShut
  {NoStop}%
\bibitem [{\citenamefont {Tews}\ \emph {et~al.}(2013)\citenamefont {Tews},
  \citenamefont {Krüger}, \citenamefont {Hebeler},\ and\ \citenamefont
  {Schwenk}}]{Tews:2012fj}%
  \BibitemOpen
  \bibfield  {author} {\bibinfo {author} {\bibfnamefont {I.}~\bibnamefont
  {Tews}}, \bibinfo {author} {\bibfnamefont {T.}~\bibnamefont {Krüger}},
  \bibinfo {author} {\bibfnamefont {K.}~\bibnamefont {Hebeler}},\ and\ \bibinfo
  {author} {\bibfnamefont {A.}~\bibnamefont {Schwenk}},\ }\href
  {https://doi.org/10.1103/PhysRevLett.110.032504} {\bibfield  {journal}
  {\bibinfo  {journal} {Phys.Rev.Lett.}\ }\textbf {\bibinfo {volume} {110}},\
  \bibinfo {pages} {032504} (\bibinfo {year} {2013})}\BibitemShut {NoStop}%
\bibitem [{\citenamefont {Lattimer}\ and\ \citenamefont
  {Lim}(2013)}]{Lattimer:2012xj}%
  \BibitemOpen
  \bibfield  {author} {\bibinfo {author} {\bibfnamefont {J.~M.}\ \bibnamefont
  {Lattimer}}\ and\ \bibinfo {author} {\bibfnamefont {Y.}~\bibnamefont {Lim}},\
  }\href {https://doi.org/10.1088/0004-637X/771/1/51} {\bibfield  {journal}
  {\bibinfo  {journal} {Astrophys.J.}\ }\textbf {\bibinfo {volume} {771}},\
  \bibinfo {pages} {51} (\bibinfo {year} {2013})}\BibitemShut {NoStop}%
\bibitem [{\citenamefont {Tews}\ \emph {et~al.}(2017)\citenamefont {Tews},
  \citenamefont {Lattimer}, \citenamefont {Ohnishi},\ and\ \citenamefont
  {Kolomeitsev}}]{Kolomeitsev:2016sjl}%
  \BibitemOpen
  \bibfield  {author} {\bibinfo {author} {\bibfnamefont {I.}~\bibnamefont
  {Tews}}, \bibinfo {author} {\bibfnamefont {J.~M.}\ \bibnamefont {Lattimer}},
  \bibinfo {author} {\bibfnamefont {A.}~\bibnamefont {Ohnishi}},\ and\ \bibinfo
  {author} {\bibfnamefont {E.~E.}\ \bibnamefont {Kolomeitsev}},\ }\href
  {https://doi.org/10.3847/1538-4357/aa8db9} {\bibfield  {journal} {\bibinfo
  {journal} {Astrophys.J.}\ }\textbf {\bibinfo {volume} {848}},\ \bibinfo
  {pages} {105} (\bibinfo {year} {2017})}\BibitemShut {NoStop}%
\bibitem [{\citenamefont {Cromartie}\ \emph {et~al.}(2020)\citenamefont
  {Cromartie}, \citenamefont {Fonseca}, \citenamefont {Ransom}, \citenamefont
  {Demorest}, \citenamefont {Arzoumanian}, \citenamefont {Blumer},
  \citenamefont {Brook}, \citenamefont {DeCesar}, \citenamefont {Dolch},
  \citenamefont {Ellis}, \citenamefont {Ferdman}, \citenamefont {Ferrara},
  \citenamefont {Garver-Daniels}, \citenamefont {Gentile}, \citenamefont
  {Jones}, \citenamefont {Lam}, \citenamefont {Lorimer}, \citenamefont {Lynch},
  \citenamefont {McLaughlin}, \citenamefont {Ng}, \citenamefont {Nice},
  \citenamefont {Pennucci}, \citenamefont {Spiewak}, \citenamefont {Stairs},
  \citenamefont {Stovall}, \citenamefont {Swiggum},\ and\ \citenamefont
  {Zhu}}]{Cromartie:2019kug}%
  \BibitemOpen
  \bibfield  {author} {\bibinfo {author} {\bibfnamefont {H.~T.}\ \bibnamefont
  {Cromartie}}, \bibinfo {author} {\bibfnamefont {E.}~\bibnamefont {Fonseca}},
  \bibinfo {author} {\bibfnamefont {S.~M.}\ \bibnamefont {Ransom}}, \bibinfo
  {author} {\bibfnamefont {P.~B.}\ \bibnamefont {Demorest}}, \bibinfo {author}
  {\bibfnamefont {Z.}~\bibnamefont {Arzoumanian}}, \bibinfo {author}
  {\bibfnamefont {H.}~\bibnamefont {Blumer}}, \bibinfo {author} {\bibfnamefont
  {P.~R.}\ \bibnamefont {Brook}}, \bibinfo {author} {\bibfnamefont {M.~E.}\
  \bibnamefont {DeCesar}}, \bibinfo {author} {\bibfnamefont {T.}~\bibnamefont
  {Dolch}}, \bibinfo {author} {\bibfnamefont {J.~A.}\ \bibnamefont {Ellis}},
  \bibinfo {author} {\bibfnamefont {R.~D.}\ \bibnamefont {Ferdman}}, \bibinfo
  {author} {\bibfnamefont {E.~C.}\ \bibnamefont {Ferrara}}, \bibinfo {author}
  {\bibfnamefont {N.}~\bibnamefont {Garver-Daniels}}, \bibinfo {author}
  {\bibfnamefont {P.~A.}\ \bibnamefont {Gentile}}, \bibinfo {author}
  {\bibfnamefont {M.~L.}\ \bibnamefont {Jones}}, \bibinfo {author}
  {\bibfnamefont {M.~T.}\ \bibnamefont {Lam}}, \bibinfo {author} {\bibfnamefont
  {D.~R.}\ \bibnamefont {Lorimer}}, \bibinfo {author} {\bibfnamefont {R.~S.}\
  \bibnamefont {Lynch}}, \bibinfo {author} {\bibfnamefont {M.~A.}\ \bibnamefont
  {McLaughlin}}, \bibinfo {author} {\bibfnamefont {C.}~\bibnamefont {Ng}},
  \bibinfo {author} {\bibfnamefont {D.~J.}\ \bibnamefont {Nice}}, \bibinfo
  {author} {\bibfnamefont {T.~T.}\ \bibnamefont {Pennucci}}, \bibinfo {author}
  {\bibfnamefont {R.}~\bibnamefont {Spiewak}}, \bibinfo {author} {\bibfnamefont
  {I.~H.}\ \bibnamefont {Stairs}}, \bibinfo {author} {\bibfnamefont
  {K.}~\bibnamefont {Stovall}}, \bibinfo {author} {\bibfnamefont {J.~K.}\
  \bibnamefont {Swiggum}},\ and\ \bibinfo {author} {\bibfnamefont {W.~W.}\
  \bibnamefont {Zhu}},\ }\href {https://doi.org/10.1038/s41550-019-0880-2}
  {\bibfield  {journal} {\bibinfo  {journal} {Nature Astronomy}\ }\textbf
  {\bibinfo {volume} {4}},\ \bibinfo {pages} {72} (\bibinfo {year}
  {2020})}\BibitemShut {NoStop}%
\bibitem [{\citenamefont {Borsanyi}\ \emph {et~al.}(2014)\citenamefont
  {Borsanyi}, \citenamefont {Fodor}, \citenamefont {Hoelbling}, \citenamefont
  {Katz}, \citenamefont {Krieg},\ and\ \citenamefont
  {Szabo}}]{Borsanyi:2013bia}%
  \BibitemOpen
  \bibfield  {author} {\bibinfo {author} {\bibfnamefont {S.}~\bibnamefont
  {Borsanyi}}, \bibinfo {author} {\bibfnamefont {Z.}~\bibnamefont {Fodor}},
  \bibinfo {author} {\bibfnamefont {C.}~\bibnamefont {Hoelbling}}, \bibinfo
  {author} {\bibfnamefont {S.~D.}\ \bibnamefont {Katz}}, \bibinfo {author}
  {\bibfnamefont {S.}~\bibnamefont {Krieg}},\ and\ \bibinfo {author}
  {\bibfnamefont {K.~K.}\ \bibnamefont {Szabo}},\ }\href
  {https://doi.org/10.1016/j.physletb.2014.01.007} {\bibfield  {journal}
  {\bibinfo  {journal} {Phys.Lett.}\ }\textbf {\bibinfo {volume} {B730}},\
  \bibinfo {pages} {99} (\bibinfo {year} {2014})}\BibitemShut {NoStop}%
\bibitem [{\citenamefont {Bazavov}\ \emph {et~al.}(2014)\citenamefont
  {Bazavov}, \citenamefont {Bhattacharya}, \citenamefont {DeTar}, \citenamefont
  {Ding}, \citenamefont {Gottlieb}, \citenamefont {Gupta}, \citenamefont
  {Hegde}, \citenamefont {Heller}, \citenamefont {Karsch}, \citenamefont
  {Laermann}, \citenamefont {Levkova}, \citenamefont {Mukherjee}, \citenamefont
  {Petreczky}, \citenamefont {Schmidt}, \citenamefont {Schroeder},
  \citenamefont {Soltz}, \citenamefont {Soeldner}, \citenamefont {Sugar},
  \citenamefont {Wagner},\ and\ \citenamefont {Vranas}}]{Bazavov:2014pvz}%
  \BibitemOpen
  \bibfield  {author} {\bibinfo {author} {\bibfnamefont {A.}~\bibnamefont
  {Bazavov}}, \bibinfo {author} {\bibfnamefont {T.}~\bibnamefont
  {Bhattacharya}}, \bibinfo {author} {\bibfnamefont {C.}~\bibnamefont {DeTar}},
  \bibinfo {author} {\bibfnamefont {H.-T.}\ \bibnamefont {Ding}}, \bibinfo
  {author} {\bibfnamefont {S.}~\bibnamefont {Gottlieb}}, \bibinfo {author}
  {\bibfnamefont {R.}~\bibnamefont {Gupta}}, \bibinfo {author} {\bibfnamefont
  {P.}~\bibnamefont {Hegde}}, \bibinfo {author} {\bibfnamefont {U.~M.}\
  \bibnamefont {Heller}}, \bibinfo {author} {\bibfnamefont {F.}~\bibnamefont
  {Karsch}}, \bibinfo {author} {\bibfnamefont {E.}~\bibnamefont {Laermann}},
  \bibinfo {author} {\bibfnamefont {L.}~\bibnamefont {Levkova}}, \bibinfo
  {author} {\bibfnamefont {S.}~\bibnamefont {Mukherjee}}, \bibinfo {author}
  {\bibfnamefont {P.}~\bibnamefont {Petreczky}}, \bibinfo {author}
  {\bibfnamefont {C.}~\bibnamefont {Schmidt}}, \bibinfo {author} {\bibfnamefont
  {C.}~\bibnamefont {Schroeder}}, \bibinfo {author} {\bibfnamefont {R.~A.}\
  \bibnamefont {Soltz}}, \bibinfo {author} {\bibfnamefont {W.}~\bibnamefont
  {Soeldner}}, \bibinfo {author} {\bibfnamefont {R.}~\bibnamefont {Sugar}},
  \bibinfo {author} {\bibfnamefont {M.}~\bibnamefont {Wagner}},\ and\ \bibinfo
  {author} {\bibfnamefont {P.}~\bibnamefont {Vranas}},\ }\href
  {https://doi.org/10.1103/PhysRevD.90.094503} {\bibfield  {journal} {\bibinfo
  {journal} {Physical Review D}\ }\textbf {\bibinfo {volume} {90}},\ \bibinfo
  {pages} {094503} (\bibinfo {year} {2014})},\ \bibinfo {note} {arXiv:
  1407.6387}\BibitemShut {NoStop}%
\bibitem [{\citenamefont {Raaijmakers}\ \emph {et~al.}(2019)\citenamefont
  {Raaijmakers}, \citenamefont {Riley}, \citenamefont {Watts}, \citenamefont
  {Greif}, \citenamefont {Morsink}, \citenamefont {Hebeler}, \citenamefont
  {Schwenk}, \citenamefont {Hinderer}, \citenamefont {Nissanke}, \citenamefont
  {Guillot}, \citenamefont {Arzoumanian}, \citenamefont {Bogdanov},
  \citenamefont {Gendreau}, \citenamefont {Ho}, \citenamefont {Lattimer},
  \citenamefont {Ludlam},\ and\ \citenamefont {Wolff}}]{Raaijmakers:2019qny}%
  \BibitemOpen
  \bibfield  {author} {\bibinfo {author} {\bibfnamefont {G.}~\bibnamefont
  {Raaijmakers}}, \bibinfo {author} {\bibfnamefont {T.~E.}\ \bibnamefont
  {Riley}}, \bibinfo {author} {\bibfnamefont {A.~L.}\ \bibnamefont {Watts}},
  \bibinfo {author} {\bibfnamefont {S.~K.}\ \bibnamefont {Greif}}, \bibinfo
  {author} {\bibfnamefont {S.~M.}\ \bibnamefont {Morsink}}, \bibinfo {author}
  {\bibfnamefont {K.}~\bibnamefont {Hebeler}}, \bibinfo {author} {\bibfnamefont
  {A.}~\bibnamefont {Schwenk}}, \bibinfo {author} {\bibfnamefont
  {T.}~\bibnamefont {Hinderer}}, \bibinfo {author} {\bibfnamefont
  {S.}~\bibnamefont {Nissanke}}, \bibinfo {author} {\bibfnamefont
  {S.}~\bibnamefont {Guillot}}, \bibinfo {author} {\bibfnamefont
  {Z.}~\bibnamefont {Arzoumanian}}, \bibinfo {author} {\bibfnamefont
  {S.}~\bibnamefont {Bogdanov}}, \bibinfo {author} {\bibfnamefont {D.~C.
  K.~C.}\ \bibnamefont {Gendreau}}, \bibinfo {author} {\bibfnamefont
  {W.~C.~G.}\ \bibnamefont {Ho}}, \bibinfo {author} {\bibfnamefont {J.~M.}\
  \bibnamefont {Lattimer}}, \bibinfo {author} {\bibfnamefont {R.~M.}\
  \bibnamefont {Ludlam}},\ and\ \bibinfo {author} {\bibfnamefont {M.~T.}\
  \bibnamefont {Wolff}},\ }\href {https://doi.org/10.3847/2041-8213/ab451a}
  {\bibfield  {journal} {\bibinfo  {journal} {The Astrophysical Journal}\
  }\textbf {\bibinfo {volume} {887}},\ \bibinfo {pages} {L22} (\bibinfo {year}
  {2019})}\BibitemShut {NoStop}%
\bibitem [{\citenamefont {Möller}\ \emph {et~al.}(2017)\citenamefont
  {Möller}, \citenamefont {Sierk}, \citenamefont {Ichikawa},\ and\
  \citenamefont {Sagawa}}]{Moller:2015fba}%
  \BibitemOpen
  \bibfield  {author} {\bibinfo {author} {\bibfnamefont {P.}~\bibnamefont
  {Möller}}, \bibinfo {author} {\bibfnamefont {A.~J.}\ \bibnamefont {Sierk}},
  \bibinfo {author} {\bibfnamefont {T.}~\bibnamefont {Ichikawa}},\ and\
  \bibinfo {author} {\bibfnamefont {H.}~\bibnamefont {Sagawa}},\ }\href
  {https://doi.org/10.1016/j.adt.2015.10.002} {\bibfield  {journal} {\bibinfo
  {journal} {Atom.Data Nucl.Data Tabl.}\ }\textbf {\bibinfo {volume}
  {109-110}},\ \bibinfo {pages} {1} (\bibinfo {year} {2017})}\BibitemShut
  {NoStop}%
\bibitem [{\citenamefont {Audi}\ \emph {et~al.}(2017)\citenamefont {Audi},
  \citenamefont {Kondev}, \citenamefont {Wang}, \citenamefont {Huang},\ and\
  \citenamefont {Naimi}}]{Audi:2017asy}%
  \BibitemOpen
  \bibfield  {author} {\bibinfo {author} {\bibfnamefont {G.}~\bibnamefont
  {Audi}}, \bibinfo {author} {\bibfnamefont {F.}~\bibnamefont {Kondev}},
  \bibinfo {author} {\bibfnamefont {M.}~\bibnamefont {Wang}}, \bibinfo {author}
  {\bibfnamefont {W.}~\bibnamefont {Huang}},\ and\ \bibinfo {author}
  {\bibfnamefont {S.}~\bibnamefont {Naimi}},\ }\href
  {https://doi.org/10.1088/1674-1137/41/3/030001} {\bibfield  {journal}
  {\bibinfo  {journal} {Chinese Physics C}\ }\textbf {\bibinfo {volume} {41}},\
  \bibinfo {pages} {030001} (\bibinfo {year} {2017})}\BibitemShut {NoStop}%
\bibitem [{\citenamefont {Ropke}\ \emph {et~al.}(2013)\citenamefont {Ropke},
  \citenamefont {Bastian}, \citenamefont {Blaschke}, \citenamefont {Klahn},
  \citenamefont {Typel},\ and\ \citenamefont {Wolter}}]{Ropke:2012qv}%
  \BibitemOpen
  \bibfield  {author} {\bibinfo {author} {\bibfnamefont {G.}~\bibnamefont
  {Ropke}}, \bibinfo {author} {\bibfnamefont {N.-U.}\ \bibnamefont {Bastian}},
  \bibinfo {author} {\bibfnamefont {D.}~\bibnamefont {Blaschke}}, \bibinfo
  {author} {\bibfnamefont {T.}~\bibnamefont {Klahn}}, \bibinfo {author}
  {\bibfnamefont {S.}~\bibnamefont {Typel}},\ and\ \bibinfo {author}
  {\bibfnamefont {H.~H.}\ \bibnamefont {Wolter}},\ }\href
  {https://doi.org/10.1016/j.nuclphysa.2012.10.005} {\bibfield  {journal}
  {\bibinfo  {journal} {Nucl.Phys.}\ }\textbf {\bibinfo {volume} {A897}},\
  \bibinfo {pages} {70} (\bibinfo {year} {2013})}\BibitemShut {NoStop}%
\bibitem [{\citenamefont {Röpke}(2015)}]{Ropke:2014fia}%
  \BibitemOpen
  \bibfield  {author} {\bibinfo {author} {\bibfnamefont {G.}~\bibnamefont
  {Röpke}},\ }\href {https://doi.org/10.1103/PhysRevC.92.054001} {\bibfield
  {journal} {\bibinfo  {journal} {Phys.Rev.}\ }\textbf {\bibinfo {volume}
  {C92}},\ \bibinfo {pages} {054001} (\bibinfo {year} {2015})}\BibitemShut
  {NoStop}%
\bibitem [{\citenamefont {Röpke}(2020)}]{Ropke:2020peo}%
  \BibitemOpen
  \bibfield  {author} {\bibinfo {author} {\bibfnamefont {G.}~\bibnamefont
  {Röpke}},\ }\href {https://doi.org/10.1103/PhysRevC.101.064310} {\bibfield
  {journal} {\bibinfo  {journal} {Physical Review C}\ }\textbf {\bibinfo
  {volume} {101}},\ \bibinfo {pages} {064310} (\bibinfo {year} {2020})},\
  \bibinfo {note} {arXiv: 2004.09773}\BibitemShut {NoStop}%
\bibitem [{\citenamefont {Typel}(2018)}]{Typel:2018wmm}%
  \BibitemOpen
  \bibfield  {author} {\bibinfo {author} {\bibfnamefont {S.}~\bibnamefont
  {Typel}},\ }\href {https://doi.org/10.1088/1361-6471/aadea5} {\bibfield
  {journal} {\bibinfo  {journal} {Journal of Physics G: Nuclear and Particle
  Physics}\ }\textbf {\bibinfo {volume} {45}},\ \bibinfo {pages} {114001}
  (\bibinfo {year} {2018})},\ \bibinfo {note} {publisher: IOP
  Publishing}\BibitemShut {NoStop}%
\bibitem [{\citenamefont {Fischer}\ \emph
  {et~al.}(2020{\natexlab{b}})\citenamefont {Fischer}, \citenamefont {Typel},
  \citenamefont {Röpke}, \citenamefont {Bastian},\ and\ \citenamefont
  {Martínez-Pinedo}}]{Fischer:2020krf}%
  \BibitemOpen
  \bibfield  {author} {\bibinfo {author} {\bibfnamefont {T.}~\bibnamefont
  {Fischer}}, \bibinfo {author} {\bibfnamefont {S.}~\bibnamefont {Typel}},
  \bibinfo {author} {\bibfnamefont {G.}~\bibnamefont {Röpke}}, \bibinfo
  {author} {\bibfnamefont {N.-U.~F.}\ \bibnamefont {Bastian}},\ and\ \bibinfo
  {author} {\bibfnamefont {G.}~\bibnamefont {Martínez-Pinedo}},\ }\href
  {https://doi.org/10.1103/PhysRevC.102.055807} {\bibfield  {journal} {\bibinfo
   {journal} {Physical Review C}\ }\textbf {\bibinfo {volume} {102}},\ \bibinfo
  {pages} {055807} (\bibinfo {year} {2020}{\natexlab{b}})},\ \bibinfo {note}
  {publisher: American Physical Society arXiv: 2008.13608}\BibitemShut
  {NoStop}%
\bibitem [{\citenamefont {Bastian}\ and\ \citenamefont
  {Blaschke}(2016)}]{Bastian:2015avq}%
  \BibitemOpen
  \bibfield  {author} {\bibinfo {author} {\bibfnamefont {N.-U.}\ \bibnamefont
  {Bastian}}\ and\ \bibinfo {author} {\bibfnamefont {D.}~\bibnamefont
  {Blaschke}},\ }\href {https://doi.org/10.1088/1742-6596/668/1/012042}
  {\bibfield  {journal} {\bibinfo  {journal} {Journal of Physics: Conference
  Series}\ }\textbf {\bibinfo {volume} {668}},\ \bibinfo {pages} {012042}
  (\bibinfo {year} {2016})}\BibitemShut {NoStop}%
\bibitem [{\citenamefont {Marczenko}\ \emph {et~al.}(2020)\citenamefont
  {Marczenko}, \citenamefont {Blaschke}, \citenamefont {Redlich},\ and\
  \citenamefont {Sasaki}}]{Marczenko:2020jma}%
  \BibitemOpen
  \bibfield  {author} {\bibinfo {author} {\bibfnamefont {M.}~\bibnamefont
  {Marczenko}}, \bibinfo {author} {\bibfnamefont {D.}~\bibnamefont {Blaschke}},
  \bibinfo {author} {\bibfnamefont {K.}~\bibnamefont {Redlich}},\ and\ \bibinfo
  {author} {\bibfnamefont {C.}~\bibnamefont {Sasaki}},\ }\href
  {http://arxiv.org/abs/2004.09566} {\bibfield  {journal} {\bibinfo  {journal}
  {arXiv:2004.09566 [astro-ph, physics:hep-ph, physics:nucl-th]}\ } (\bibinfo
  {year} {2020})},\ \bibinfo {note} {arXiv: 2004.09566}\BibitemShut {NoStop}%
\bibitem [{\citenamefont {Blaschke}\ and\ \citenamefont
  {Ropke}(1988)}]{Blaschke:1988zt}%
  \BibitemOpen
  \bibfield  {author} {\bibinfo {author} {\bibfnamefont {D.}~\bibnamefont
  {Blaschke}}\ and\ \bibinfo {author} {\bibfnamefont {G.}~\bibnamefont
  {Ropke}},\ }\href@noop {} {\bibfield  {journal} {\bibinfo  {journal}
  {Z.Phys.A}\ } (\bibinfo {year} {1988})}\ \bibinfo {note} {Preprint: JINR-E2-88-77}\BibitemShut {NoStop}%
\bibitem [{\citenamefont {Typel}(2016)}]{Typel:2016srf}%
  \BibitemOpen
  \bibfield  {author} {\bibinfo {author} {\bibfnamefont {S.}~\bibnamefont
  {Typel}},\ }\href {https://doi.org/10.1140/epja/i2016-16016-3} {\bibfield
  {journal} {\bibinfo  {journal} {Eur.Phys.J.}\ }\textbf {\bibinfo {volume}
  {A52}},\ \bibinfo {pages} {16} (\bibinfo {year} {2016})}\BibitemShut
  {NoStop}%
\bibitem [{\citenamefont {Bastian}\ \emph {et~al.}(2018)\citenamefont
  {Bastian}, \citenamefont {Blaschke}, \citenamefont {Fischer},\ and\
  \citenamefont {Röpke}}]{Bastian:2018wfl}%
  \BibitemOpen
  \bibfield  {author} {\bibinfo {author} {\bibfnamefont {N.-U.~F.}\
  \bibnamefont {Bastian}}, \bibinfo {author} {\bibfnamefont {D.}~\bibnamefont
  {Blaschke}}, \bibinfo {author} {\bibfnamefont {T.}~\bibnamefont {Fischer}},\
  and\ \bibinfo {author} {\bibfnamefont {G.}~\bibnamefont {Röpke}},\ }\href
  {https://doi.org/10.3390/universe4060067} {\bibfield  {journal} {\bibinfo
  {journal} {Universe}\ }\textbf {\bibinfo {volume} {4}},\ \bibinfo {pages}
  {67} (\bibinfo {year} {2018})}\BibitemShut {NoStop}%
\bibitem [{\citenamefont {Bastian}\ and\ \citenamefont
  {Blaschke}(2018)}]{Bastian:2018mmc}%
  \BibitemOpen
  \bibfield  {author} {\bibinfo {author} {\bibfnamefont {N.-U.~F.}\
  \bibnamefont {Bastian}}\ and\ \bibinfo {author} {\bibfnamefont {D.~B.}\
  \bibnamefont {Blaschke}},\ }\href@noop {} {\bibfield  {journal} {\bibinfo
  {journal} {arXiv:1812:11766 [nucl-th]}\ } (\bibinfo {year} {2018})},\
  \bibinfo {note} {arXiv: 1812.11766 [nucl-th]}\BibitemShut {NoStop}%
\bibitem [{\citenamefont {Typel}\ and\ \citenamefont
  {Blaschke}(2018)}]{Typel:2017vif}%
  \BibitemOpen
  \bibfield  {author} {\bibinfo {author} {\bibfnamefont {S.}~\bibnamefont
  {Typel}}\ and\ \bibinfo {author} {\bibfnamefont {D.}~\bibnamefont
  {Blaschke}},\ }\href {https://doi.org/10.3390/universe4020032} {\bibfield
  {journal} {\bibinfo  {journal} {Universe}\ }\textbf {\bibinfo {volume} {4}},\
  \bibinfo {pages} {32} (\bibinfo {year} {2018})}\BibitemShut {NoStop}%
\bibitem [{\citenamefont {Roark}\ and\ \citenamefont
  {Dexheimer}(2018)}]{Roark:2018uls}%
  \BibitemOpen
  \bibfield  {author} {\bibinfo {author} {\bibfnamefont {J.}~\bibnamefont
  {Roark}}\ and\ \bibinfo {author} {\bibfnamefont {V.}~\bibnamefont
  {Dexheimer}},\ }\href {https://doi.org/10.1103/PhysRevC.98.055805} {\bibfield
   {journal} {\bibinfo  {journal} {Phys.Rev.}\ }\textbf {\bibinfo {volume}
  {C98}},\ \bibinfo {pages} {055805} (\bibinfo {year} {2018})}\BibitemShut
  {NoStop}%
\bibitem [{\citenamefont {Sagert}\ \emph {et~al.}(2009)\citenamefont {Sagert},
  \citenamefont {Fischer}, \citenamefont {Hempel}, \citenamefont {Pagliara},
  \citenamefont {Schaffner-Bielich}, \citenamefont {Mezzacappa}, \citenamefont
  {Thielemann},\ and\ \citenamefont {Liebendörfer}}]{Sagert:2008ka}%
  \BibitemOpen
  \bibfield  {author} {\bibinfo {author} {\bibfnamefont {I.}~\bibnamefont
  {Sagert}}, \bibinfo {author} {\bibfnamefont {T.}~\bibnamefont {Fischer}},
  \bibinfo {author} {\bibfnamefont {M.}~\bibnamefont {Hempel}}, \bibinfo
  {author} {\bibfnamefont {G.}~\bibnamefont {Pagliara}}, \bibinfo {author}
  {\bibfnamefont {J.}~\bibnamefont {Schaffner-Bielich}}, \bibinfo {author}
  {\bibfnamefont {A.}~\bibnamefont {Mezzacappa}}, \bibinfo {author}
  {\bibfnamefont {F.-K.}\ \bibnamefont {Thielemann}},\ and\ \bibinfo {author}
  {\bibfnamefont {M.}~\bibnamefont {Liebendörfer}},\ }\href
  {https://doi.org/10.1103/PhysRevLett.102.081101} {\bibfield  {journal}
  {\bibinfo  {journal} {Physical Review Letters}\ }\textbf {\bibinfo {volume}
  {102}},\ \bibinfo {pages} {081101} (\bibinfo {year} {2009})},\ \bibinfo
  {note} {publisher: American Physical Society}\BibitemShut {NoStop}%
\bibitem [{\citenamefont {Marques}\ \emph {et~al.}(2017)\citenamefont
  {Marques}, \citenamefont {Oertel}, \citenamefont {Hempel},\ and\
  \citenamefont {Novak}}]{Marques:2017zju}%
  \BibitemOpen
  \bibfield  {author} {\bibinfo {author} {\bibfnamefont {M.}~\bibnamefont
  {Marques}}, \bibinfo {author} {\bibfnamefont {M.}~\bibnamefont {Oertel}},
  \bibinfo {author} {\bibfnamefont {M.}~\bibnamefont {Hempel}},\ and\ \bibinfo
  {author} {\bibfnamefont {J.}~\bibnamefont {Novak}},\ }\href
  {https://doi.org/10.1103/PhysRevC.96.045806} {\bibfield  {journal} {\bibinfo
  {journal} {Physical Review C}\ }\textbf {\bibinfo {volume} {96}},\ \bibinfo
  {pages} {045806} (\bibinfo {year} {2017})},\ \bibinfo {note} {arXiv:
  1706.02913}\BibitemShut {NoStop}%
\bibitem [{\citenamefont {Baym}\ \emph {et~al.}(2019)\citenamefont {Baym},
  \citenamefont {Furusawa}, \citenamefont {Hatsuda}, \citenamefont {Kojo},\
  and\ \citenamefont {Togashi}}]{Baym:2019iky}%
  \BibitemOpen
  \bibfield  {author} {\bibinfo {author} {\bibfnamefont {G.}~\bibnamefont
  {Baym}}, \bibinfo {author} {\bibfnamefont {S.}~\bibnamefont {Furusawa}},
  \bibinfo {author} {\bibfnamefont {T.}~\bibnamefont {Hatsuda}}, \bibinfo
  {author} {\bibfnamefont {T.}~\bibnamefont {Kojo}},\ and\ \bibinfo {author}
  {\bibfnamefont {H.}~\bibnamefont {Togashi}},\ }\href
  {https://doi.org/10.3847/1538-4357/ab441e} {\bibfield  {journal} {\bibinfo
  {journal} {The Astrophysical Journal}\ }\textbf {\bibinfo {volume} {885}},\
  \bibinfo {pages} {42} (\bibinfo {year} {2019})},\ \bibinfo {note} {arXiv:
  1903.08963}\BibitemShut {NoStop}%
\end{thebibliography}%

\end{document}